\documentstyle[12pt,aasms4,psfig]{article}

\lefthead{K. Krisciunas {\em et al.}}
\righthead{Observations through SDSS Filters}

\begin{document}

\title{The Recognition of Unusual Objects in the 
Sloan Digital Sky Survey Color System} 
\author{Kevin Krisciunas, Bruce Margon, and Paula Szkody}
\affil{Department of Astronomy, University of Washington, Box 351580,
Seattle, WA 98195$-$1580}
\begin {center}
Electronic mail: kevin, margon, szkody@astro.washington.edu
\end {center}

\begin{abstract}
We present 5 filter photometry of 
21 carbon stars, 15 asteroids, 15 cataclysmic variables,
6 metal-poor stars, 5 Cepheids, 1775 field stars, blue
horizontal branch (BHB) stars and RR Lyrae stars in  the globular clusters
M 15 and M 2, two primary standards, and 19 secondary standards.  
The photometry was carried out using a filter set identical to that 
which will be used for the Sloan Digital Sky Survey.  We find that
carbon stars, CVs, R-type, J-type, and V-type
asteroids, BHB stars, and RR Lyr stars should be
identifiable on the basis of SDSS photometry alone, while Cepheids, 
metal-poor stars, and many types of asteroids are indistinguishable from
the stellar locus of field stars.

\end{abstract}

\keywords{Sloan Digital Sky Survey, photometry, carbon stars,
asteroids, cataclysmic variables, Cepheids, RR Lyrae stars,
metal-poor stars, horizontal branch stars}

\section{Introduction}

The Sloan Digital Sky Survey (SDSS) will image 10$^4$ deg$^2$ of
the northern high-latitude sky to approximately magnitude 23 in five
special filter bands from roughly 3500 to 9000 \AA \hspace {1 mm}
(\markcite{gk93}Gunn \& Knapp 1993; \markcite{gw95}Gunn \& Weinberg 1995; 
\markcite{m98}Margon 1998). The broad-band filters of the SDSS system
cover the full wavelength range of an optical CCD.  The bandpasses overlap
very little, thus allowing essentially unique portions of an object's
spectral energy distribution to be sampled by each filter.  Furthermore,
sky features such as the 5460 \AA \hspace {1 mm} Hg I line, the 5577 \AA
\hspace {1 mm} line of [O I] and a strong OH band fall between the
bandpasses, thus allowing more accurate photometry because one is not so
much at the mercy of short-term changes of the sky brightness.  The
resulting magnitudes will be on the AB system, referred to a spectrum with
constant $f_{\nu}$ (rather than an A0 V star), and represent the fluxes a
flat spectrum source with a given $V$ magnitude would have near the
effective wavelengths of the filters. 

The resulting SDSS photometry
of roughly 10$^8$ objects (stars, galaxies, quasars, etc.)
will be used to select targets for multifiber spectroscopy, as well as
constitute a permanent archive for a large variety of astrophysical problems.
For both of these functions, it is imperative to understand whether (and
how) interesting subclasses of objects stand out compared to the normal
stellar locus in multi-color space. Further, the precise transformation
between the SDSS color system and more traditional ones will become very
important for a variety of future studies.

Although the exact SDSS photometric system requires use of the actual survey 
telescopes, cameras, and filters, it is already possible to mimic the system 
reasonably well.  The names of the Survey bandpasses are u$^{\prime}$,
g$^{\prime}$, r$^{\prime}$, i$^{\prime}$, and z$^{\prime}$,
and their effective wavelengths (\markcite{f96}Fukugita et al. 1996, 
hereafter F96) are 3557, 4825, 6161, 7672, and 9097 \AA, respectively.
While the photometric filter profiles and CCD characteristics of SDSS 
photometry have been defined, a fully calibrated set of photometric
standards is still being established.  We therefore follow the example of 
\markcite{r97}Richards et al. (1997) and refer to our observations, as
u$^*$ data, g$^*$ data, etc.  

The observations and analysis of this paper have several purposes:
(1) to quantify the effectiveness of the SDSS filters
for isolating unusual classes of objects from the normal
stellar locus; (2) to complement
previous work and further the determination of the
stellar locus of normal stars in the four Sloan colors;
\footnote[1]{In addition to the observational papers cited above
on this problem, \markcite{l98}Lenz et al. (1998) present extensive
synthetic photometry which is also relevant.} and
(3) to determine the magnitudes and colors of a list of potential
secondary standard stars, which will aid further SDSS observations
and contribute to the establishment of the Sloan system.

\markcite {r97}Richards et al. (1997) and \markcite {new97}Newberg 
et al. (1997) have used the SDSS Monitor Telescope at Apache Point Observatory
(New Mexico) and the 1-m telescope of the United States Naval 
Observatory (USNO) to delineate the locus of field stars and the
positions of quasars in SDSS multi-color space.
Our observations here on a larger number of fields (but typically to
shallower photometric depth) confirm the position of the stellar locus.
We present observations of carbon stars, cataclysmic variables (CVs),
asteroids, some metal-poor stars, and some globular cluster stars of
interest (Cepheids, RR Lyrae stars, and horizontal branch stars).
While the exact photometric values of our observations will change once
the official SDSS calibration is established, we believe that the
observations presented here will require relatively small corrections.
The {\em qualitative} results should be unchanged, since
objects that separate from the stellar locus {\em now} will still separate
from the stellar locus, no matter how the ``sausage'' is twisted in
multi-color space by the final calibration.

\section{Observations and Data Reduction}

\subsection{Techniques}

Our observations were obtained using the USNO 1-m 
Ritchey-Chr$\acute{\rm e}$tien reflector in
Flagstaff, Arizona.  The camera contains an LN$_2$ cooled
1024 $\times$ 1024 charge-coupled device (CCD).  The gain is
$\approx$7 electrons per analog-to-digital unit (ADU), with a read noise
of 6 electrons.  The plate scale is 0.68 arcsec pixel$^{-1}$,
giving a field of view of 11.5 $\times$ 11.5 arcmin.  
Since non-linear effects begin at $\approx$16,000 ADUs, we
occasionally had to obtain extra, shorter, exposures on
some bright stars, or defocus the camera slightly if the
required exposure would have to have been less than 4 seconds.

The USNO camera contains a UV-enhanced CCD nearly identical to 
the u$^{\prime}$ CCDs on the SDSS 2.5-m telescope. 
The {\em other} four bands of the camera on the 2.5-m 
telescope do not use UV-enhanced chips.
The effective wavelengths for the other four filters
used for the observations presented in this paper
are thus slightly different than the values given by F96.
The smaller SDSS monitor telescope will also use a single
UV-enhanced CCD to define the SDSS system.

One of the major drawbacks of the USNO 1-m telescope,
which was originally completed in 1934, is that one
typically cannot accurately guide the telescope
when pointed at an object further than two hours from the meridian.
Thus, to observe standards at a wide range of air mass requires that 
some of the standards be rather far south.  On some nights we did observe
some stars twice, at air masses $\approx$1.2 and $\approx$1.8, 
thus allowing zeroth order estimates of the extinction in each filter.
We did not obtain sufficient data on red-blue pairs of stars to derive
the {\em second} order atmospheric coefficients according
to the standard method described by \markcite {h62}Hardie (1962).  Thus
we performed only first order atmospheric corrections.

Altogether we observed on 9 nights from UT dates 27 September to
5 October 1997.  Nights 2, 3, 4, 8, and 9
were photometric from start to finish.  Half of nights 1, 5, and 7
were good, as was the end of night 6.  We took care to use only
data bracketed by consistent standards, indicating which portions
of the partial nights were indeed perfectly clear.
We observed the SDSS primary
standard BD +17$^{\rm o}$ 4708 on nights 1, 3, 4, 8, and 9.  
BD +21$^{\rm o}$ 607, which is intended to become a primary standard,
was observed on nights 2, 7, and 8.  

In addition to the primary standards, we observed 19 secondary
standards from a list drawn up for the Sloan project (Allyn Smith,
private communication).  These are mostly standards of 
\markcite {l92}Landolt (1992) -- stars known to be constant 
in brightness and with known Johnson UBV and Cousins RI magnitudes.  
On full nights we obtained between 14 and 19 sets of 5 filter standards 
data.  

Except for the globular cluster data, we reduced all our data
with the {\sc iraf} data reduction package.  This first
involved subtracting the bias frames, constructing sky flats with the
median of three sky frames  per filter per night, then flattening and
trimming the frames.  For the
data on stars and asteroids we determined the instrumental
magnitudes as aperture magnitudes, using {\bf phot}
within the {\bf apphot} package.  We almost always used an
aperture radius of 6 pixels. This was after the inspection of
many stars in many frames using {\bf imexam}.  For sets of
frames we selected the field stars to be reduced using {\bf daofind}.
Our selection criterion was 10 $\sigma$ over sky in
the u$^*$ filter.  This was because we wanted to maximize 
the number of stars detected in all 5 filters, and the
u$^*$ filter plus the near-UV quantum efficiency of the CCD gives 
the least efficient response.
Calibration of the data was done within the {\bf photcal} package,
in particular with {\bf mknobsfile}, {\bf fitparams},
and {\bf evalfit}.

The derivation of a table of standard star magnitudes and colors
was a time consuming, iterative, but convergent process.  The
values in A. Smith's list of Sloan secondary
standards were generated for the purpose
of estimating exposure times, not for the purpose of data
reduction.  But given the hour angle constraints of the telescope
mentioned above, we needed {\em some} set of values for the derivation
of atmospheric extinction and atmospheric reddening if we could not
observe
the same stars at a range of air mass.  We therefore temporarily
adopted the predicted r$^*$ magnitudes and four colors for the
stars and adjusted these values until: (1) we obtained sensibly
linear photometric solutions; and (2) the {\sc rms} internal
errors of the fits on photometric nights were smaller than
0.02 mag.  By ``sensible linear photometric solution'' we 
mean the following.  We obtained instrumental magnitudes in
each of the 5 SDSS filters.  To obtain r$^*$ magnitudes let us
designate the instrumental r$^*$ and i$^*$ magnitudes by 
r$_{instr}$ and i$_{instr}$, the mean air mass over the course of 
the r$^*$ exposure by $\overline{X_r}$, and the first order
r$^*$-band extinction by $k_r$.  We have:

\begin{displaymath}
r^* \; = \; ZP_r \; + r_{instr} \; - \; k_r \; \overline{X_r} \; + \; 
c_r \; (r_{instr} \; - i_{instr}) \; , 
\end{displaymath}

\parindent = 0 mm

where ZP$_r$ is the zero point and c$_r$ is a scaling coefficient.
Of course, we could have just as easily used the instrumental
g$^*-$r$^*$ color in the previous equation.  For the
reduction of r$^*-$i$^*$ colors we have an equation such as:

\begin{displaymath}
r^* - i^* \; = \; ZP_{ri} \; - \; k_{ri} \; \overline {X_{ri}} 
\; + \; c_{ri} \; (r_{instr} \; - i_{instr}) \; , 
\end{displaymath}

where ZP$_{ri}$ is the zero point, $k_{ri}$ is the first order
atmospheric reddening, and
$c_{ri}$ is a scaling coefficient.  There are, of course,
analogous expressions for the other three SDSS colors.
If we were to include second order terms, in the first equation
we would use a slightly different first order extinction term, say 
$k^{\prime}_r$, and we would have to add a term such as 
$k^{\prime \prime}_r \;  \overline {X_{ri}} \; (r_{instr} - i_{instr})$.
The r$^*-$i$^*$ color reduction equation would require, instead, 
a first order term such as $k^{\prime}_{ri}$ and 
would require the addition of a term such as
$k^{\prime \prime}_{ri} \; \overline {X_{ri}} \; (r_{instr} - i_{instr})$.
There would be analogous expressions for other SDSS colors.
While F96 gives estimates for the second order terms, we decided to
set them to zero.  F96 suggests that the largest second order
term is for u$^*-$g$^*$ reductions for stars with u$^*-$g$^*$
bluer than +1.3.  For those stars the inclusion of a second order 
term would add at most a few hundredths of a magnitude to the
scatter of the data.  For stars with u$^*-$g$^*$ redder than +1.3,
such as our carbon stars, some of which are very red indeed,
F96 suggests that the second order color term for u$^*-$g$^*$
reductions is exactly 0.00, in which case we may be able
to take at face value the derived u$^*-$g$^*$ colors of the reddest
stars presented here.

\parindent = 9 mm

We discovered that it is possible to construct a table of
secondary standards
that gives sensible extinction and atmospheric reddening, low {\sc
rms} internal errors for the nightly fits, and which gives the right
values for BD +17$^{\rm o}$ 4708, but which does {\em not}
give c$_r$ = 0.00, or c$_{ug}$, c$_{gr}$, c$_{ri}$, c$_{iz}$ = 1.00.
This would be a photometric ``system'', but not a system  that would
allow the Sloan magnitudes to be converted to AB magnitudes 
for stars with colors {\em other than} those close to the colors 
of BD +17$^{\rm o}$ 4708.

Because the SDSS standard star network is still incomplete, 
conventional photometric reduction procedures, including the
use of color terms, are not appropriate.
For those wholly photometric nights when BD +17$^{\rm o}$ 4708 was 
observed, we determined the ``standardized values'' for
the secondary standards by means of single filter differential
photometry with respect to this primary, using first order
extinction corrections, then adding the differential values to the
values of primary (which are known from spectrophotometry).
This gave a new standards table, which we then tested with {\sc iraf}.
The nightly solutions gave nearly the same extinction and atmospheric
reddening values as those used to derive the magnitudes and colors of
the secondary stars.  Within {\sc iraf} the resulting color
coefficients used to scale instrumental magnitudes and colors
were, within very small errors, equal to 0.00 for the r$^*$-band data, 
and 1.00 for the four colors.   Arithmetically,
this is the same as simply correcting the instrumental magnitudes (or
colors) of the program objects and field stars to
outside-the-atmosphere values with first order atmospheric terms.

We derived the zero points and first order extinction and
atmospheric reddening
parameters using as much information as the 9 night observing 
run provided -- as many as two primary and 
17 secondary standards per night.  This is
better than deriving the nightly zero points solely on the basis
of one star, no matter how primary it is. It could have been the
case that we were observing through thin cirrus only at the time we
were measuring the primary standard.  

As an example of the success of our photometric method, we show in
Fig. 1 a plot of the u$^*-$g$^*$ photometric solution for the night of 29 
September 1997, using a first order atmospheric reddening term of
$k_{ug} =  0.26$ mag/air mass.
The $y$-intercept is the photometric zero point
for this night for this color index; it is not zero because of the
different bandpass widths of the u$^*$ and g$^*$ filters and the
variation of quantum effficiency of the chip as a function of wavelength.
The slope of the line is statistically indistinguishable from
 1.00, and the {\sc RMS}
residual of the fit is a remarkably low $\pm 0.010$ mag. Other
photometric solutions (i.~e. different colors and different nights)
are very similar to Fig. 1, if appropriate atmospheric terms are used.
This means that our primary standard and secondary standards 
are self-consistent, and we believe will closely approximate the
eventual, official SDSS system.

Based on 6 nights of observations at the USNO Flagstaff site,
on which we observed an adequate number of standards under clear
skies and over a wide enough air mass range, we
find a mean first order r$^*$-band extinction of $k_r$ = 0.109 $\pm$ 0.018
mag/air mass.  We find mean atmospheric reddening values (in the same
units)
of $k_{ug}$ = 0.292 $\pm$ 0.043, $k_{gr}$ = 0.083 $\pm$ 0.008,
$k_{ri}$ = 0.067 $\pm$ 0.017, and $k_{iz}$ = $-0.013 \pm 0.018$.
\footnote[2]
{$k_{iz} < 0$ implies that the z$^*$-band extinction
was greater than the i$^*$-band extinction.  Since near-infrared
extinction is dominated by water vapor emission, not scattering
by aerosols, it is physically reasonable that $k_{iz}$ can be
negative.  See Fig. 1 of \markcite{k87}Krisciunas et al. (1987).}
For comparison, F96 give predictions for Apache Point of $k_r$ =
0.092, $k_{ug} \approx$ 0.38, $k_{gr}$ = 0.090, $k_{ri}$ = 0.013, and
$k_{iz}$ = $-$0.032 mag/air mass.

In Table 1 we give r$^*$ magnitudes and four
colors of two primary standards and 19 secondary standards,
with the values rounded off to the nearest hundredth.  In most
cases the measurements match, to better than a hundredth, those derived
filter by filter on the four whole nights when BD +17$^{\rm o}$ 4708 
was observed. In some cases we have adjusted the values by 0.01 or 
0.02 mag by taking into account observations made over the course
of the whole observing run. More observations of these stars 
would certainly be welcome, and redder standards must
be added to the list.

F96 predicts transformations between the Johnson UBV
and Cousins RI systems and the SDSS system.  We find
similar, but not identical, transformations from actual 
observations of real stars, and suggest that the following
equations be used:

\begin{displaymath}
r^* \; = \; V \; - \; 0.42 \; (B-V) \; + \; 0.10 \; ,
\end{displaymath}

\begin{displaymath}
u^* \; - g^* \; = \; 1.33 \; (U-B) \; + \; 1.18 \; ,
\end{displaymath}

\begin{displaymath}
g^* \; - r^* \; = \; 0.96 \; (B-V) \; - \; 0.18 \; ,
\end{displaymath}

\begin{displaymath}
r^* \; - i^* \; = \; 0.99 \; (R-I) \; - \; 0.22 \; ,
\end{displaymath}

\begin{displaymath}
i^* \; - z^* \; = \; 0.67 \; (R-I) \; - \; 0.19 \; .
\end{displaymath}

\parindent = 0 mm

F96 gives two transformations between R$-$I and
r$^*-$i$^*$, depending on whether R$-$I is less than
or greater than +1.15.  Since our reddest standard has 
R$-$I = 0.81, we can only consider one of the transformations.
These transformations will probably change yet again when the final
SDSS photometry calibration is determined.  For now it
seems more sensible to use our transformations, based
on actual data, rather than use the predicted transformations
given in F96.

\parindent = 9 mm

In Tables 2, 3, and 4 we give photometric results for cataclysmic variables,
carbon stars, and asteroids, respectively.  For the most part we do
not state the internal errors of the points, as they are often less than
$\pm$0.01 mag, and very rarely larger than $\pm$0.02 mag.

\subsection{Cataclysmic Variables}

Since CVs occupy some of the same color phase space as quasars, we wish
to see how well SDSS filters will differentiate them from each other.
In Table 2 we give the CV types, the orbital periods, and the nightly
means
of r* and four colors for 15 CVs. The CVs were selected to cover a range of
orbital periods, magnetic field
strengths, and mass transfer rates, all of which 
affect the colors. CVs which have long orbital periods, or which are in
low mass transfer states, usually show the secondary star, so they will be
redder. Large magnetic fields can result in red cyclotron emission, and large
mass transfer rates lead to steady state disks which dominate the emission,
yielding a flux distribution that is proportional to $\lambda^{-2.3}$
(\markcite{w95}Warner 1995).  

Because the dwarf novae undergo outbursts, and nova-like systems undergo
high and low states of mass transfer, it is fortuitous that
the objects were observed at
various states.  The dwarf novae AR And, V1159 Ori, and RX And were
near outburst.  (RX And was observed on four nights as it cooled
and faded toward quiescence.)  The disk nova-like system TT Ari was in a
high state while  the magnetic system AM Her was in a low state.
 
\subsection{Carbon Stars}

It is one of many stellar astronomy goals of SDSS systematically to
discover significant numbers of faint high latitude carbon stars
(FHLCs) (\markcite{ti98}Totten \& Irwin 1998), which are rare 
but interesting objects for multiple reasons.  The most extreme examples 
(\markcite{bm84}Margon et al. 1984)
are at galactocentric distances of up to 100 kpc, and thus almost
surely well beyond the dark matter halo. Such objects are thus crucial
dynamical probes, and their sharp bandheads make accurate spectroscopic
velocity determinations straightforward.  Some uncertain but probably
small fraction of FHLCs are in fact nearby dwarfs (\markcite{gm94}Green 
\& Margon 1994, \markcite{g98}Green 1998). Although only a dozen 
such objects are currently known, all are within $\sim$100~pc, and 
thus it seems certain that the
until-recently unrecognized class of dC stars are by a large factor the
numerically most common carbon stars in the Galaxy, overwhelming the 
well-studied C giants.

We selected more than 20 FHLC stars for observation, primarily from the
lists of {\markcite{b91}Bothun et al. (1991), \markcite{sp88}Sanduleak 
\& Pesch (1988), and \markcite{skb69}Slettebak, Keenan \& Brundage (1969). 
Unfortunately only two dC stars were available during this observing 
run, although they have the virtue of including
two of the brightest known objects, including the prototype, viz.
G77-61 and LHS 1075.

In Table 3 we give our averaged data for 19 ``regular'' carbon stars 
and two dwarf carbon stars.  Many of these stars are probably variable 
in brightness to some degree.  For example, according the 4th edition 
of the {\em General
Catalogue of Variable Stars} (\markcite{gcvs}Kholopov 1987), AM Scl 
irregularly varies between magnitude 12.42 and 13.29.
For those few stars in Table 3 which were observed on more than one night,
we see no evidence of variability.  

\subsection{Asteroids}

We are interested to know if asteroids of different taxonomic types
(\markcite{tb89}Tholen \& Barucci 1989) separate from the stellar locus in
SDSS multi-color space; or, will SDSS photometry at least allow us
to differentiate different types of asteroids from each other?

We observed 15 asteroids of 5 taxonomic types.  In Table 4 we give
nightly averages of the photometry.  Three of the
asteroids were observed on more than one night.  Because asteroids
can be variable in brightness and color, we do not average the data
from more than one night.

The taxonomic types given in Table 4 are from \markcite{x95}Xu et al.
(1995), with two exceptions.  For 371 Bohemia, Xu et al. give its
type as S.  This object is now classed AS, meaning ``olivine-rich
S-type with near infrared absorption band minimum typically falling
past 1 micron'' (Hammergren, personal communication).  The near-Earth
asteroid 433 Eros is an S-type asteroid (\markcite{l96}Lee et al. 1996).

\subsection{Metal-Poor Stars and Cluster Variables}

In Table 5 we give photometry of six metal-poor stars, four 
Cepheids in M 2, and one Cepheid in M 15.  
The globular cluster data were reduced using Dophot version 2.0 (\markcite 
{dophot}Schechter, Mateo, \& Saha 1993), along with some
notes from the Canary Islands Winter School (\markcite{m96}Mateo 1996).
This code uses point spread function (PSF) fitting, a must when trying to
reduce data in crowded stellar fields.
From aperture photometry, reduced with {\sc iraf},
of stars in more empty parts of the M2 and M 15 frames, we were able to
put the Dophot results on the same photometric system as our standard stars,
with estimated systematic errors of 0.03 mag or less.  

\section{Discussion}

In Figures 2, 3, and 4 we show our program objects (CVs, carbon
stars, asteroids, metal-poor stars, and Cepheids) in color-color
space.  In Figures 5, 6, and 7 we add the locus of field stars.

Let $\sigma_a$ and $\sigma_b$ be the internal errors
of the magnitudes of a star in filters $a$ and $b$, respectively,
and let $\sigma_{ab}$ = $\surd (\sigma_{a}^{2} + 
\sigma_{b}^{2})$.  In our color-color plots the field stars are
only plotted if the corresponding value of $\sigma_{ab}$ is less
than 0.10 mag.  While this reduces the number of stars in our
field star sample, it sharpens the edges of the field star locus
in multi-color space.

The stellar locus was derived from roughly 80 fields at a wide variety
of galactic latitudes, and amounts to 1775 stars selected on the
basis of their instrumental u$^*$ counts to be at least 10 $\sigma$ 
times the sky noise, with the criterion on internal errors mentioned
above.    In Fig. 5 we have included
quasars observed by \markcite{r97}Richards et al. (1997), which have
redshifts ranging 0.06 $< z < $ 2.7. 

Objects shown in Figures 2 through 4 that do not appear
in Figures 5 through 7 do not separate themselves from the
stellar locus and thus can not be identified on the basis of
their SDSS colors only.  This includes metal-poor stars,
Cepheids, and most asteroids.

However, as one can see from the figures, the CVs easily separate 
from the stellar locus, especially in the bluest colors.
This is not that surprising, since a CV consists typically 
of a hot star, an accretion disk, and a cool star.  The
reddest objects are the longer period systems GK Per,
SS Cyg, AE Aqr, and SS Aur, as well as the magnetic system
AM Her in a low state.  In all of these the red color is
due to the contribution of the secondary star.  At the same
time the addition of a blue accretion disk/column separates
them from the field stars.  Most CVs occupy the region near
(0,0) in all Sloan color-color plots, which is also the area 
occupied by low redshift quasars (see Fig. 5) and white dwarfs 
(Fig. 4 of \markcite{l98}Lenz et al. 1998).  Thus we expect
new CVs and white dwarfs to be by-products of the quasar
discoveries.

While in Figs. 5 and 6 all the points corresponding to RX And
stand out from the stellar locus, in
Fig. 7 we see that RX And at outburst is separable
from the general stellar locus, but as the system
returns to a quiescent state, its colors blend with the stellar
locus in the i$^*-$z$^*$ vs. r$^*-$i$^*$ diagram.

Fig. 6 clearly shows that carbon stars separate from the stellar
locus in r$^*-$i$^*$ vs. g$^*-$r$^*$, more so at the reddest
g$^*-$r$^*$ colors.  This in entirely in accord with synthetic
photometry carried out by P. J. Green (private communication).
Clearly, we need to observe more known dwarf carbon stars to see
if they all have colors like G77-61 and LHS~1075, the colors of
which are curiously identical within the errors.  In any case, SDSS photometry
will be effective for identifying carbon stars in the halo,
although probably not for separating dwarfs from giants.

Since one of the primary goals of the Sloan project is to identify
a very large number of
quasar candidates, we naturally were interested in whether we
managed to image any quasars serendipitously.  As shown in Fig. 5,
these would be objects that stand out from the stellar locus
in the u$^*-$g$^*$ vs. g$^*-$r$^*$ diagram.  
In Table 6 we give photometry of 5 objects with u$^*$ 
excesses.  One of them, in fact, {\em is} a previously catalogued
quasar.  Near CM~Del we imaged 2022+171 (= MG 2024+1717 =
J202456.5+171814), a quasar of $z$ = 1.05.  Interestingly,
it is at the same location in Fig. 5 as PG 2302+029, a quasar
also of redshift 1.05, observed by \markcite{r97}Richards et al.
(1997).  Another blue object we serendipitously recovered is PB 7559
(\markcite{bf84}Berger \& Fringant 1984).  To our knowledge 
there is no published spectrum; it could be a star or a quasar.
The other three objects in our Table 6 are most
likely stars, given their brightness.  

Of the asteroids we observed, only one, 349 Dembowska, separates from
the stellar locus (see Figs. 7 and 10).  This is an R-type asteroid,
notable for its {\em lack} of near infrared light, thus giving an
i$^*-$z$^*$ color that is ``too blue''.  From the spectra given by
\markcite{x95}Xu et al. (1995), we expect that J-type and V-type asteroids
will also stand out for the same reason.

In Figures 8, 9, and 10 we show photometry of the asteroids only.  
For reference, the colors of the Sun are also shown.  Clearly, the C-type
asteroids separate themselves from the other types of asteroids observed.
This will allow us to identify C-type asteroids from SDSS photometry
of moving objects.\footnote[3] {We note that C-type asteroids are further
subdivided into types B, F, and G depending on the near-ultraviolet 
light (\markcite{tb89}Tholen \& Barucci 1989, Fig. 5; \markcite{x95}Xu et
al. 1995).  The u$^*-$g$^*$ colors of a large sample of C-type
asteroids may allow further differentiation using SDSS data.}

In Fig. 11 we show a color-magnitude diagram of the globular cluster M 15,
where we have only plotted those points with Dophot internal errors
(for each filter) less than 0.1 mag.
M 15 has clearly identifiable blue horizontal branch (BHB) stars, located
in Fig. 11 in the pentagonal box.  The region of the RR Lyr stars 
would be just to
the right.  Since the RR Lyr stars are being observed at random phases
in their pulsation cycles, we would expect these HB stars
in the instability strip to show a wider spread in r$^*$ than the
BHB stars, and in fact they do show this.

It is worth commenting that the broadening of the lower giant branch
is undoubtedly an artifact of crowding.  As one approaches
the frame limit, it becomes harder and harder to separate stars,
especially in a globular cluster field.  Therefore, if the counts
from a given pixel in the frame are from two stars of comparable magnitude
but widely varying colors, a ``star'' found by Dophot could be
up to 0.75 mag ``too bright'', with a color that is considerably
different
than the colors of either star.  And the photometry of this ``star'' could
have an internal error less than our arbitrary limit of 0.1 mag.

We have  identified 80 RR Lyr stars in our frames of M~15 using
the list of coordinates given by \markcite{hsh}Sawyer Hogg (1973).  
These stars have a median r$^*$ magnitude of 15.76.  The
standard deviation of the distribution is $\sigma_{r^*} = \pm 0.20$.
\markcite{ss95}Silbermann \& Smith (1995) obtain a mean 
V magnitude for these stars of 15.82, adopt a mean absolute 
magnitude of $<$M$_V>$ = +0.36 $\pm$ 0.12, 
color excess E(B-V) = 0.08 mag, absorption A$_V$ = 0.26 mag, 
and therefore log D(pc) = 4.04.  Assuming an interstellar absorption
law of the form A$_\lambda \propto \lambda^{-1}$, we find A$_{r^*}$ = 0.23
mag and $<$M$_{r^*}>$ = +0.33.  (This could be slightly biased toward
a value that is too bright because some of the stars may be blended
with other stars.)  Our data imply, however, that
whatever the true mean absolute visual 
magnitude of the RR Lyr stars in M~15 (a matter of some
debate), the mean absolute SDSS r$^*$ magnitude of these stars 
is only slightly brighter.

We find that the BHB and RR Lyr stars in M 15 overlap the
blue regions of the {\em entire} stellar locus in u$^*-$g$^*$ vs.
g$^*-$r$^*$ and in i$^*-$z$^*$ vs. r$^*-$i$^*$, but
in Fig. 12 we show that they are found on a track parallel to our
stellar locus, and on average 0.2 mag redder in r$^*-$i$^*$
(or, equivalently, bluer in g$^*-$r$^*$)
in the r$^*-$i$^*$ vs. g$^*-$r$^*$ diagram.  (A comparison of
Figs. 5 and 12 shows that this is also where quasars and CVs
are to be found.)  Even if some of this
displacement in r$^*-$i$^*$ is attributable to an incorrect
zero point in placing our Dophot output on our SDSS system,
since so few field stars are this blue, it means that the
identification of field BHB and RR Lyr stars from SDSS photometry
is promising.

In Table 7 we give the median SDSS r$^*$ magnitudes and
colors of our sample of 143 BHB stars
and 80 RR Lyr stars in M 15.  We also include the median 
values for 247 BHB stars and 7 RR Lyr stars in M 2.
We would expect that field BHB and RR Lyr stars found
by SDSS in uncrowded fields would have colors within 0.2
mag of the values listed in Table 7 if their metallicites
are comparable to those of the two clusters.

To investigate further the ability of SDSS to identify field RR Lyr
stars in the northern Galactic cap, we have divided our data into three
roughly equal area regions
with galactic latitudes $0 < |b| < 19^o$, 
$19^o < |b| < 42^o$ and $42^o < |b| < 90^o$.
We would expect the effects of interstellar reddening
to affect the u$^*-$g$^*$ vs. g$^*-$r$^*$ stellar locus the most.
In Fig. 13 we show that RR Lyr stars have colors that separate themselves
from the general stellar locus {\em at higher galactic latitudes}
(see lower set of points in the figure), where SDSS data will 
actually be obtained.  Since the BHB stars have
bluer g$^*-$r colors than RR Lyr stars, they are even further separated
from the stellar locus at high galactic latitudes in the u$^*-$g$^*$ vs.
g$^*-$r$^*$ diagram.  Thus, we have confidence that SDSS will be able to
identify many new BHB and RR Lyr candidates, and that the low latitude
field stars present in this study, but not in SDSS, have not confused
our conclusions.

We can use the information presented in Fig. 13 to investigate whether
we can demonstrate any effects of interstellar reddening on SDSS colors.
The upper set of 858 points (the stars at lower galactic latitudes)
shows a wider locus of points than the lower set of 272 points. Is
this just the effect of having a larger data set, putting a larger
{\em number} of stars further from the mid-range, or do the
distributions have different standard deviations?
We begin by fitting an orthogonal regression line (also known as 
``Pearson's major axis'') to each of the two stellar loci
(\markcite{i90}Isobe et al. 1990).  These are the lines that minimize the
squares of the {\em perpendiculars} from the points to the lines.  
The two slopes are the same within the errors.  The stars at lower
galactic latitudes are displaced, on average, 0.039 mag to redder
u$^*-$g$^*$ colors, and are displaced 0.018 mag to redder g$^*-$r$^*$
colors. These displacements, however, are not statistically significantly
different than zero, given the uncertainties in the u$^*-$g$^*$ intercepts
of the plots.  

If we now calculate, for each set of points, the {\sc rms} deviation from
the respective orthogonal regression line (i.e. perpendicular to the
line), we find for the lower latitude sample 
$\sigma_{\perp} = \pm 0.096$ mag,
while for the higher latitude sample 
$\sigma_{\perp} = \pm 0.066$ mag.  The
uncertainties of these standard deviations are $\pm$ 0.002 and $\pm$
0.003, respectively.  \footnote[4] {The reader may recall that one
can obtain an unbiased estimate of the variance ($\hat{\sigma}^2$)
of a distribution of points.  An unbiased estimate of the variance of the
variance is 2($\hat{\sigma}^2)^2/(n-1$), where $n$ is the sample size
(\markcite{b65}Brownlee 1965).  It can be shown that the {\em fractional}
uncertainty of the {\sc rms} error of a distribution is 
$\frac{1}{\surd 2(n-1)}$.}  Thus, the {\em widths} of the distributions 
differ by roughly 10 times their uncertainties.  
The mid-latitude stars with  $19^o < |b| < 42^o$ give $\sigma_{\perp} =
\pm 0.081$ mag.  

Another way of looking at the stellar locus according to ranges of
galactic latitude is to pose the null hypothesis that the widths of the
distributions shown in Fig. 13 are the same.  Using a two sided F-test, we
reject this null hypothesis at a 99.9 percent confidence level.  Our data
demonstrate that the stellar locus {\em is} widened by some combination of
interstellar reddening and metallicity as we proceed from higher galactic
latitudes to lower galactic latitudes. 

\section{Conclusions}

Using data provided by the Fermilab group for standards in common
and graphs of the stellar locus (Newberg 1998, private communication), we
have confirmed the position of the stellar locus in SDSS multi-color 
space found previously by \markcite{r97}Richards et al. (1997)
and \markcite{new97}Newberg et al. (1997).
 Using our stellar locus, we have shown that cataclysmic variables,
carbon stars, and R-type asteroids separate from the stellar
locus on the basis of their SDSS colors.  J-type asteroids, V-type
asteroids, blue horizontal branch stars and
RR Lyr stars should also be identifiable  from SDSS photometry.  
C-type asteroids are easily distinguishable from asteroids of type
A, S, AS, and R on the basis of SDSS colors.  However, Cepheids,
metal-poor stars, and many types of asteroids do not separate from the
stellar locus in color-color space, and thus cannot be identified on the
basis of SDSS photometry alone.

We also provide a list of secondary stars that, after the completion
of the SDSS photometric system, may be useful for calibrating
future photometry.  More observations of these stars should be
obtained, and redder stars should also be observed.

\acknowledgments

KK thanks Hugh Harris of the US Naval Observatory for arranging
the telescope time at the 1-m telescope and for observing
assistance and advice; Alan Diercks, Scott Anderson, and 
Guillermo Gonzalez for data reduction advice; Doug Duncan
for a list of metal-poor stars; George Wallerstein for information
on globular clusters; Mark Hammergren for providing a list of
coordinates of asteroids spanning the time of our observing run;
and Heidi Newberg,
Brian Yanny, Gordon Richards, and Steve Kent for useful
discussions.  Chris Stubbs provided much useful discussion
and greatly appreciated moral and financial support.
Useful information was obtained from
Simbad, the astronomical data base of the Centre de
Donn$\acute{\rm e}$es astronomiques de Strasbourg.


\newpage 

\begin{deluxetable}{lrccccccc}
\tablewidth{0pc}
\tablecaption{Two Primary and 19 Secondary Standards$^a$}
\tablehead{
\colhead{Star Name} & \colhead{r$^*$} &
\colhead{u$^*-$g$^*$} & \colhead{g$^*-$r$^*$} &
\colhead{r$^*-$i$^*$} & \colhead{i$^*-$z$^*$} &
\colhead{N} & \colhead{n$_u$} & 
\colhead{n$_{griz}$} }
\startdata
BD +21$^{\rm o}$ 607   & 9.12   &  0.89  & 0.28 & 0.10 & $-$0.01 &  3 &  4 &  3 \nl
BD +17$^{\rm o}$ 4708  & 9.35   &  0.92  & 0.29 & 0.10 &  0.02 &  5 &  5 &  5 \nl
\nl
SA 92-342      & 11.52    & 1.15     & 0.25     & 0.05    & $-$0.03  & 6 & 11  & 6 \nl     
SA 92-263      & 11.45    & 2.32     & 0.80     & 0.30    & 0.14   & 6 & 11  & 6 \nl
SA 94-242      & 11.72    & 1.23     & 0.10     & $-$0.02   & $-$0.06  & 6 & 12  & 8 \nl
SA 94-251      & 10.78    & 2.83     & 0.98     & 0.37    & 0.19   & 6 & 12  & 8 \nl
Ross 374       & 10.64    & 1.05     & 0.37     & 0.16    & 0.03   & 1 &  1  & 1 \nl

SA 95-218      & 11.89    & 1.49     & 0.49     & 0.16    & 0.04   & 6 & 12  & 7 \nl
SA 95-132      & 11.98    & 1.54     & 0.26     & 0.07    & 0.03   & 6 & 12  & 7 \nl
SA 95-142      & 12.74    & 1.33     & 0.43     & 0.16    & 0.05   & 6 & 12  & 7 \nl
SA 95-149      & 10.37    & 3.30     & 1.33     & 0.61    & 0.35   & 2 &  4  & 2 \nl
BD $-21^{\rm o}$ 910     &  9.60    & 1.50     & 0.46     & 0.12    & 0.05   & 3 &  5  & 3 \nl

SA 110-499     & 11.38    & 2.11     & 0.82     & 0.46    & 0.31   & 5 &  9  & 7 \nl
SA 110-503     & 11.60    & 1.84     & 0.46     & 0.22    & 0.12   & 5 &  9  & 7 \nl
SA 110-441     & 10.99    & 1.32     & 0.37     & 0.11    & 0.03   & 2 &  6  & 4 \nl
SA 110-450     & 11.28    & 2.12     & 0.77     & 0.39    & 0.27   & 2 &  6  & 4 \nl
SA 112-805     & 12.17    & 1.24   & $-$0.09  & $-$0.14   & $-$0.10  & 6 & 11  & 7 \nl

SA 112-822     & 11.22    & 2.34     & 0.80     & 0.28    & 0.13   & 6 & 11  & 7 \nl
SA 113-259     & 11.37    & 2.78     & 0.93     & 0.31    & 0.19   & 4 &  7  & 4 \nl
SA 113-260     & 12.29    & 1.27     & 0.31     & 0.05    & 0.00   & 4 &  7  & 4 \nl
BD +28$^{\rm o}$ 4211    & 10.75  & $-$0.52  & $-$0.52  & $-$0.38  & $-$0.32  & 8 &  8  & 8 \nl
\nl
\enddata
\tablenotetext{a} {N is the number of nights on which a given star was observed.
n$_u$ is the number of u$^*$ frames obtained.  n$_{griz}$ is the number of frames
obtained through the other four SDSS filters.  The data for BD +17$^{\rm o}$~4708
are derived by \markcite{f96}Fukugita et al. (1996) from spectrophotometry.}
\end{deluxetable}

\begin{deluxetable}{lccccccccc}
\tablewidth{0pc}
\tablecaption{Observations of Cataclysmic Variables}
\tablehead{
\colhead{Star Name} & 
\colhead{Type$^a$} & \colhead{P$^b$} &
\colhead{JD$-$2450000} & 
\colhead{r$^*$} &
\colhead{u$^*-$g$^*$} & \colhead{g$^*-$r$^*$} &
\colhead{r$^*-$i$^*$} & \colhead{i$^*-$z$^*$} &
\colhead{n$^c$} }
\startdata
AR And    & DN & 3.9  & 726.8696  & 12.62  & 0.14  & $-$0.18  & $-$0.15  & $-$0.13  & 1 \nl
RX And    & DN & 5.0  & 722.8987  & 11.34  & 0.06  & $-$0.16  & $-$0.12  & $-$0.06  & 2 \nl
$-$      & $-$ & $-$  & 724.9118  & 12.07  & 0.09  & $-$0.10  & $-$0.01  & $-$0.05  & 2 \nl
$-$      & $-$ & $-$  & 725.8525  & 12.47  & 0.07  & $-$0.03  & 0.02   & 0.05   & 1 \nl
$-$      & $-$ & $-$  & 726.8532  & 13.08  & 0.09  &  0.03  & 0.17   & 0.12   & 1 \nl
\nl
V603 Aql  & NL & 3.3  & 722.6258  & 11.64  & $-$0.15 & $-$0.05  & $-$0.12  & $-$0.11  & 1 \nl
AE Aqr   &  IP & 9.9  & 721.7215  & 10.74  & 1.18  &  0.68  &  0.24  &  0.10  & 1 \nl
FO Aqr   &  IP & 4.9  & 720.7041  & 13.48  & $-$0.19 &  0.01  & $-$0.12  & $-$0.14  & 1 \nl
TT Ari   &  NL & 3.3  & 723.9725  & 11.16  & 0.05  & $-$0.21  & $-$0.20  & $-$0.15  & 2 \nl
SS Aur   &  DN & 4.4  & 718.9606  & 14.85  & $-$0.34 &  0.48  & 0.48   &  0.36  & 1 \nl
\nl
V425 Cas &  NL & 3.6  & 726.6912  & 15.24  & 0.13  &  0.02  &  $-$0.09 &  $-$0.07 & 1 \nl
SS Cyg   &  DN & 6.6  & 719.6611  & 11.55  & 0.33  &  0.61  &  0.31  &  0.14  & 1 \nl 
$-$      & $-$ & $-$  & 725.6642  & 11.50  & 0.38  &  0.56  &  0.32  &  0.20  & 1 \nl
$-$      & $-$ & $-$  & 726.6522  & 11.53  & 0.30  &  0.57  &  0.34  &  0.17  & 1 \nl
CM Del   &  NL & 3.9  & 722.6347  & 14.18  & 0.10  &  0.01  &  $-$0.01 &  $-$0.05 & 1 \nl
\nl
AM Her   &  AM & 3.1  & 719.6168  & 14.68  & 0.00  &  0.53  &  0.83  &  0.84  & 2 \nl
V1159 Ori & DN & 1.5 & 720.9754  & 13.59  & 0.25  & $-$0.15  & $-$0.17  & $-$0.13  & 1 \nl
LS Peg   &  NL  & 4.2 & 720.6713  & 11.92  & 0.04  & $-$0.03  & $-$0.16  & $-$0.15  & 1 \nl
GK Per   &  IP &  47.9 & 720.0037  & 12.44  & 0.87  &  0.82  &  0.38  &  0.25  & 1 \nl
AO Psc   &  IP &  3.6 & 721.7065  & 13.62  & $-$0.16 & $-$0.09  & $-$0.13  & $-$0.02  & 1 \nl
\nl
\enddata
\tablenotetext{a} {DN = dwarf nova; NL = nova-like; AM = polar; IP = intermediate polar.}
\tablenotetext{b} {P is the orbital period of the system in hours.}
\tablenotetext{c} {n is the number of 5 filter data sets obtained.  Some extra
u$^*$-band images were also obtained.}
\end{deluxetable}

\begin{deluxetable}{lrccccccc}
\tablewidth{0pc}
\tablecaption{Observations of Carbon Stars$^a$}
\tablehead{
\colhead{Star Name} & \colhead{r$^*$} &
\colhead{u$^*-$g$^*$} & \colhead{g$^*-$r$^*$} &
\colhead{r$^*-$i$^*$} & \colhead{i$^*-$z$^*$} &
\colhead{N} & \colhead{n$_u$} & 
\colhead{n$_{griz}$} }
\startdata
C01      & 13.18  & 2.99 $\pm$ 0.01 &  1.33 &  0.43 &  0.38  & 1 & 2 & 1 \nl  
C02      & 15.43  & 2.83 $\pm$ 0.03 &  1.12 &  0.32 &  0.29  & 2 & 4 & 4 \nl  
C03      & 15.25  & 3.07 $\pm$ 0.02 &  1.13 &  0.32 &  0.26  & 1 & 2 & 1 \nl  
C07      & 13.99  &                 &  2.48 &  0.91 &  0.55  & 1 & 0 & 1 \nl  
C08      & 14.66  & 2.80 $\pm$ 0.04 &  1.36 &  0.40 &  0.38  & 1 & 2 & 2 \nl
\nl
C09      & 14.40  & 4.70 $\pm$ 0.13 &  1.61 &  0.60 &  0.50  & 1 & 2 & 2 \nl  
C10      & 15.15  & 3.25 $\pm$ 0.02 &  1.25 &  0.38 &  0.35  & 2 & 4 & 3 \nl  
C13      & 18.10  & 1.09 $\pm$ 0.04 &  0.45 &  0.25 & $-$0.03  & 1 & 2 & 2 \nl  
C15      & 10.10  & 3.85 $\pm$ 0.01 &  1.75 &  0.59 &  0.56  & 1 & 3 & 1 \nl  
C23      & 14.71  & 2.57 $\pm$ 0.04 &  0.90 &  0.21 &  0.22  & 1 & 2 & 2 \nl  
\nl
C26      & 14.19  & 3.51 $\pm$ 0.02 &  1.35 &  0.39 &  0.41  & 2 & 2 & 2 \nl
C30      & 13.42  & 6.06 $\pm$ 0.32 &  1.78 &  0.55 &  0.48  & 1 & 1 & 1 \nl  
C31      & 13.99  &                 &  2.00 &  0.75 &  0.50  & 1 & 0 & 1 \nl 
CLS 105  & 13.96  & 2.35 $\pm$ 0.01 &  0.97 &  0.27 &  0.14  & 1 & 2 & 2 \nl  
CLS 112  & 14.63  & 1.94 $\pm$ 0.01 &  0.79 &  0.19 &  0.13  & 1 & 2 & 2 \nl  
\nl
G77-61   & 13.20  & 3.00 $\pm$ 0.01 &  1.53 &  0.40 &  0.15  & 2 & 4 & 3 \nl  
LHS 1075 & 14.34  & 3.02 $\pm$ 0.04 &  1.54 &  0.40 &  0.20  & 2 & 5 & 3 \nl  
SKB2     & 11.96  & 2.09 $\pm$ 0.01 &  0.79 &  0.19 &  0.12  & 1 & 2 & 2 \nl  
BD $-19^{\rm o}$ 132  & 10.26  & 3.75 $\pm$ 0.01 &  1.35 &  0.38 &  0.37  & 1 & 2 & 1 \nl  
AM Scl     & 12.02  & 5.74 $\pm$ 0.15 &  1.84 &  0.70 &  0.43  & 1 & 2 & 2 \nl  
\nl
BD $-19^{\rm o}$ 290  & 10.57  & 3.05 $\pm$ 0.01 &  1.10 &  0.29 &  0.32  & 1 & 2 & 1 \nl

\nl
\enddata
\tablenotetext{a} {Objects prefixed by a ``C'' are from 
\markcite {b91}Bothun et al. (1991); CLS objects are from 
\markcite {sp88}Sanduleak \& Pesch (1988); for G77-61 and LHS 1075
see \markcite {d94}Deutsch (1994); for the last four 
objects in the table see \markcite {skb69}Slettebak, Keenan \& Brundage (1969).
N is the number of nights on which a given star was observed.
n$_u$ is the number of u$^*$ frames obtained on which the object was detected
with an adequate S/N ratio.  n$_{griz}$ is the number of frames
obtained through the other four SDSS filters.}
\end{deluxetable}

\begin{deluxetable}{rlccccccccc}
\tablewidth{0pc}
\tablecaption{Observations of Asteroids$^a$}
\tablehead{
\colhead{No.} & \colhead{Name} & \colhead{Type} &
\colhead{1997 Date} & \colhead{UT} & 
\colhead{r$^*$} &
\colhead{u$^*-$g$^*$} & \colhead{g$^*-$r$^*$} &
\colhead{r$^*-$i$^*$} & \colhead{i$^*-$z$^*$} &
\colhead{n} }
\startdata
446 & Aeternitas & A  & 30 Sep & 11:09 & 14.33 & 1.54 & 0.61 & 0.26 & 0.12  & 1 \nl
702 & Alauda     & C  & 30 Sep & 11:47 & 12.75 & 1.58 & 0.45 & 0.13 & 0.03  & 1 \nl
82  & Alkmene    & S  & 28 Sep & 10:57 & 12.43 & 1.71 & 0.64 & 0.18 & $-$0.05 & 1 \nl
$-$  & $-$    & $-$  &  3 Oct & 10:56 & 12.27 & 1.70 & 0.60 & 0.21 & $-$0.07 & 2 \nl
774 & Armor      & S  & 28 Sep & 03:52 & 12.97 & 1.68 & 0.71 & 0.15 & 0.06  & 1 \nl
$-$ & $-$      & $-$  &  4 Oct & 03:23 & 13.21 & 1.78 & 0.65 & 0.23 & $-$0.02 & 2 \nl
\nl
371 & Bohemia    & AS &  4 Oct & 05:05 & 12.64 & 1.77 & 0.65 & 0.21 & $-$0.04 & 2 \nl
349 & Dembowska  & R  & 29 Sep & 11:54 & 10.28 & 1.92 & 0.70 & 0.15 & $-$0.22 & 1 \nl
433 & Eros$^b$   & S  & 30 Sep & 12:26 & 13.30 & 1.94 & 0.67 & 0.24 & $-$0.07 & 2 \nl
480 & Hansa      & S  & 29 Sep & 09:59 & 12.25 & 1.74 & 0.63 & 0.22 & $-$0.07 & 1 \nl
10  & Hygeia     & C  & 28 Sep & 12:05 & 11.43 & 1.56 & 0.46 & 0.10 & 0.02  & 1 \nl
\nl
683 & Lanzia     & C  &  1 Oct & 03:35 & 14.23 & 1.53 & 0.48 & 0.13 & 0.04  & 1 \nl
$-$ & $-$     & $-$  &  3 Oct & 05:01 & 14.21 & 1.62 & 0.45 & 0.14 & $-$0.02 & 1 \nl
68  & Leto       & S  & 27 Sep & 10:37 & 10.11 & 1.76 & 0.66 & 0.18 & 0.03  & 1 \nl
149 & Medusa     & S  & 29 Sep & 06:12 & 12.57 & 1.82 & 0.68 & 0.22 & $-$0.01 & 1 \nl
196 & Philomela  & S  & 28 Sep & 08:51 & 11.41 & 1.81 & 0.64 & 0.19 & $-$0.03 & 1 \nl
\nl
314 & Rosalia    & C  &  5 Oct & 03:15 & 13.95 & 1.57 & 0.46 & 0.10 &  0.00 & 2 \nl
138 & Tolosa     & S  &  5 Oct & 04:11 & 11.57 & 1.94 & 0.69 & 0.17 & $-$0.01 & 2 \nl
\nl
\enddata
\tablenotetext{a} {With one exception the Universal Time given 
corresponds to the r$^*$ image, or the mean of two r$^*$ images.  
n is the number of full 5 filter data sets obtained.}
\tablenotetext{b} {All data on Eros were interpolated to the
mean time of two 5 filter data sets.}
\end{deluxetable}

\begin{deluxetable}{lccccccc}
\tablewidth{0pc}
\tablecaption{Observations of Metal-Poor Stars and Cepheids$^a$}
\tablehead{
\colhead{Star Name} & \colhead{[Fe/H]} &
\colhead{Per(d)} & \colhead{r$^*$} &
\colhead{u$^*-$g$^*$} & \colhead{g$^*-$r$^*$} &
\colhead{r$^*-$i$^*$} & \colhead{i$^*-$z$^*$} }
\startdata
Metal-poor stars:   &         &      &      &       &       &       &      \nl
BD +72$^{\rm o}$ 94 & $-$1.80 &      & 9.83 &  0.94 &  0.26 &  0.08 &  0.02 \nl  
HD 16031            & $-$2.20 &      & 9.67 &  0.99 &  0.24 &  0.06 &  0.04 \nl
BD +3$^{\rm o}$ 740 & $-$2.90 &      & 9.68 &  0.90 &  0.30 &  0.07 &  0.02  \nl
G 186-26            & $-$2.80 &      & 10.74 &  0.85 &  0.23 &  0.07 & $-$0.01 \nl 
BD $-$17$^{\rm o}$ 6036 & $-$2.80 &      & 10.22 & 1.52 &  0.64 &  0.30 &  0.11  \nl
BD +38$^{\rm o}$ 4955   & $-$2.50 &      & 10.80 &  1.04 &  0.51 &  0.22 &  0.08 \nl
\nl
Cepheids:     &        &        &      &       &       &       \nl
M2-V1         &        & 15.583 &  12.76 & 1.70 & 0.56  & 0.09  & 0.19  \nl
M2-V5         &        & 17.606 &  13.12 & 1.92 & 0.72  & 0.23  & 0.14  \nl
M2-V6         &        & 19.295 &  12.79 & 1.11 & 0.45  & 0.11  & 0.07  \nl
M2-V11        &        & 67.0   &  11.64 & 1.35 & 0.43  & 0.11  & 0.10  \nl
M15-V1        &        & 1.438  &  15.18 & 1.34 &  0.29 &  0.17 &  0.06 \nl  
\nl
\enddata
\tablenotetext{a}{The periods of the Cepheids were taken from
\markcite{hsh}Sawyer Hogg (1973).}
\end{deluxetable}

\begin{deluxetable}{lcccccccc}
\tablewidth{0pc}
\tablecaption{Field Objects with u$^*$ Excesses}
\tablehead{
\colhead{Identification} & \colhead{$\alpha$ (2000)} &
\colhead{$\delta$ (2000)} & \colhead{b$^a$} &
\colhead{r$^*$} &
\colhead{u$^*-$g$^*$} & \colhead{g$^*-$r$^*$} &
\colhead{r$^*-$i$^*$} & \colhead{i$^*-$z$^*$} }
\startdata
          & 20:24:50.0  &   +25:00:24  & $-$7.6   & 15.11  & 0.53  & 0.02  & $-$0.10  &    0.06 \nl
2022+171  & 20:24:56.6  &   +17:18:13  & $-$12.0  & 17.81  & 0.33  & 0.20  &    0.02  &    0.01 \nl
          & 20:25:11.5  &   +25:05:45  & $-$7.6   & 15.17  & 0.40  & 0.07  & $-$0.04  &    0.03 \nl
          & 23:04:22.1  &   +53:18:45  & $-$6.1   & 16.19  & 0.40  & 0.26  &    0.13  & $-$0.01 \nl
PB 7559   & 23:07:51.1  & $-$13:40:28  & $-$62.7  & 18.03  & 0.25  & 0.03  & $-$0.02  & $-$0.11 \nl
\nl
\enddata
\tablenotetext{a} {Galactic latitude in degrees.}
\end{deluxetable}

\begin{deluxetable}{ccccccccc}
\tablewidth{0pc}
\tablecaption{Blue Horizontal Branch and RR Lyrae Stars (median values)}
\tablehead{
\colhead{Cluster} & \colhead{[Fe/H]$^a$} &
\colhead{Star Type} & \colhead{n} & 
\colhead{r$^*$} & 
\colhead{u$^*-$g$^*$} & \colhead{g$^*-$r$^*$} &
\colhead{r$^*-$i$^*$} & \colhead{i$^*-$z$^*$} }
\startdata
M 15 & $-$2.22 & BHB    & 143 & 15.99    & 1.33  & $-$0.06  & 0.02  & $-$0.08 \nl
$-$  & $-$     & RR Lyr &  80 & 15.76    & 1.27  &    0.19  & 0.16  & $-$0.02 \nl
\nl
M 2  & $-$1.62 & BHB    & 247 & 16.28    & 1.13  & $-$0.09  & $-$0.12 & $-$0.05 \nl
$-$  & $-$     & RR Lyr &   7 & 15.88    & 1.17  &    0.24  &    0.06 &    0.03 \nl
\enddata
\tablenotetext{a} {Metallicity is taken from Harris (1996).}
\end{deluxetable}

\begin{figure*}  
\psfig{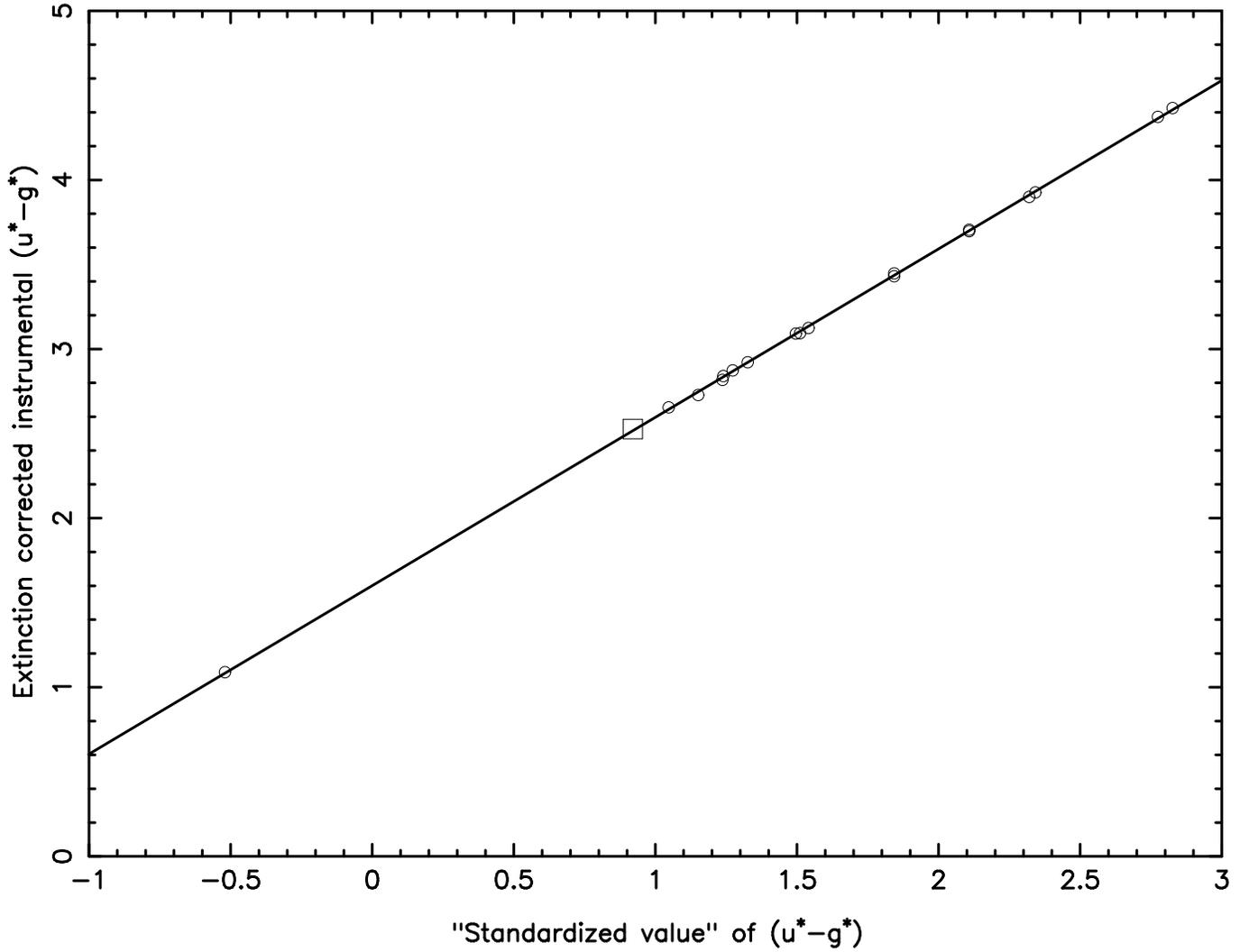}
\caption{u$^*-$g$^*$
photometric solution for 29 September 1997 UT.
The $y$-intercept is the ``photometric zero point''
for this color for this night.  The slope of the line is not
statistically different than 1.00.  The RMS
residual of the fit is $\pm$0.010 mag.
The open square represents the primary standard BD +17$^o$~4708.
The smaller open circles represent secondary standards.}
\end{figure*}

\begin{figure*}  
\psfig{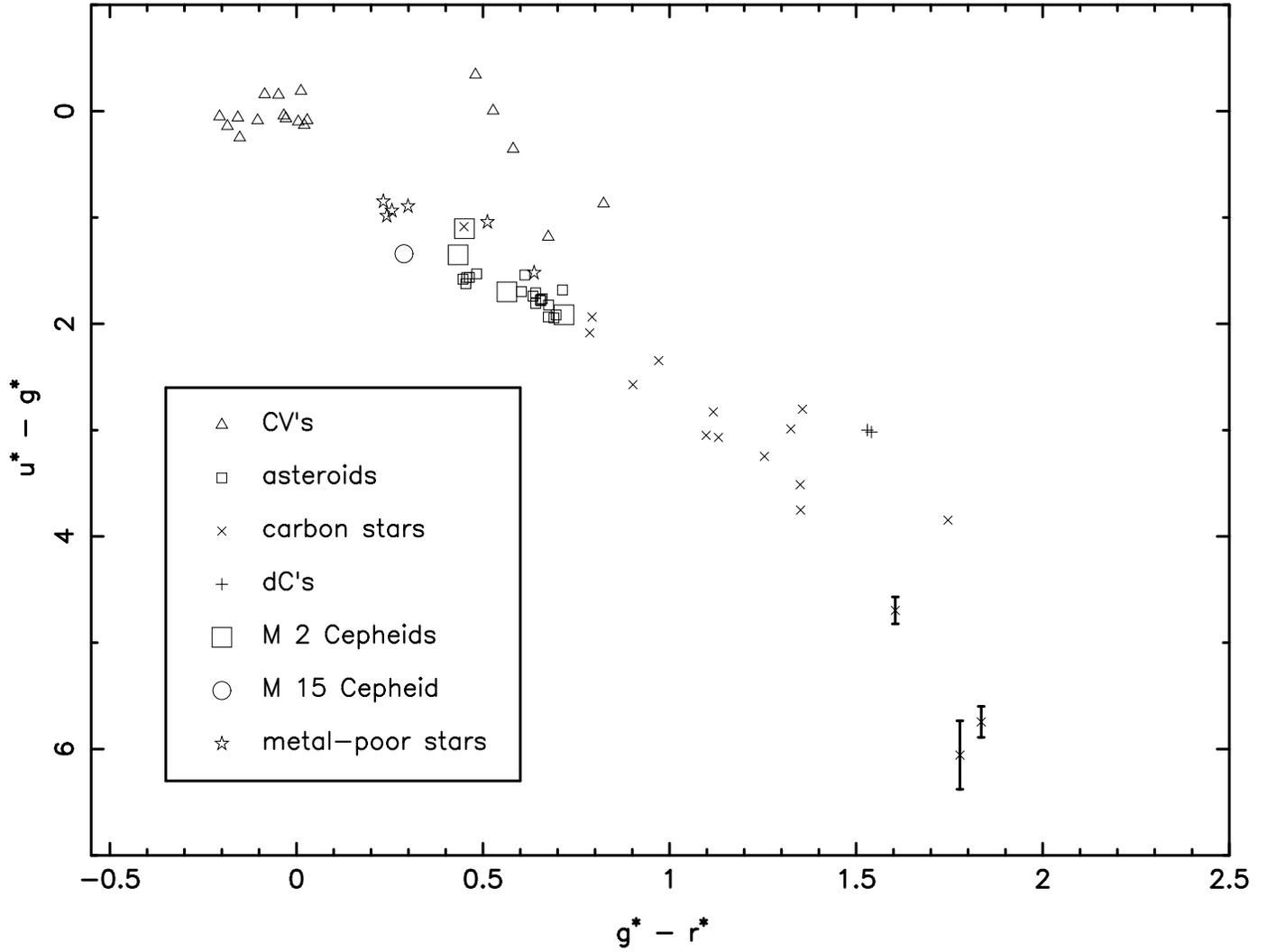}
\caption{u$^*-$g$^*$ vs. g$^*-$r$^*$
data for program objects
from Tables 2 through 5.  
The point for SS Cyg is the  average of 3 nights of data.  Some carbon
star points are the averages of 2 nights of data.  Other points
represent nightly averages.}
\end{figure*}

\begin{figure*}  
\psfig{figure=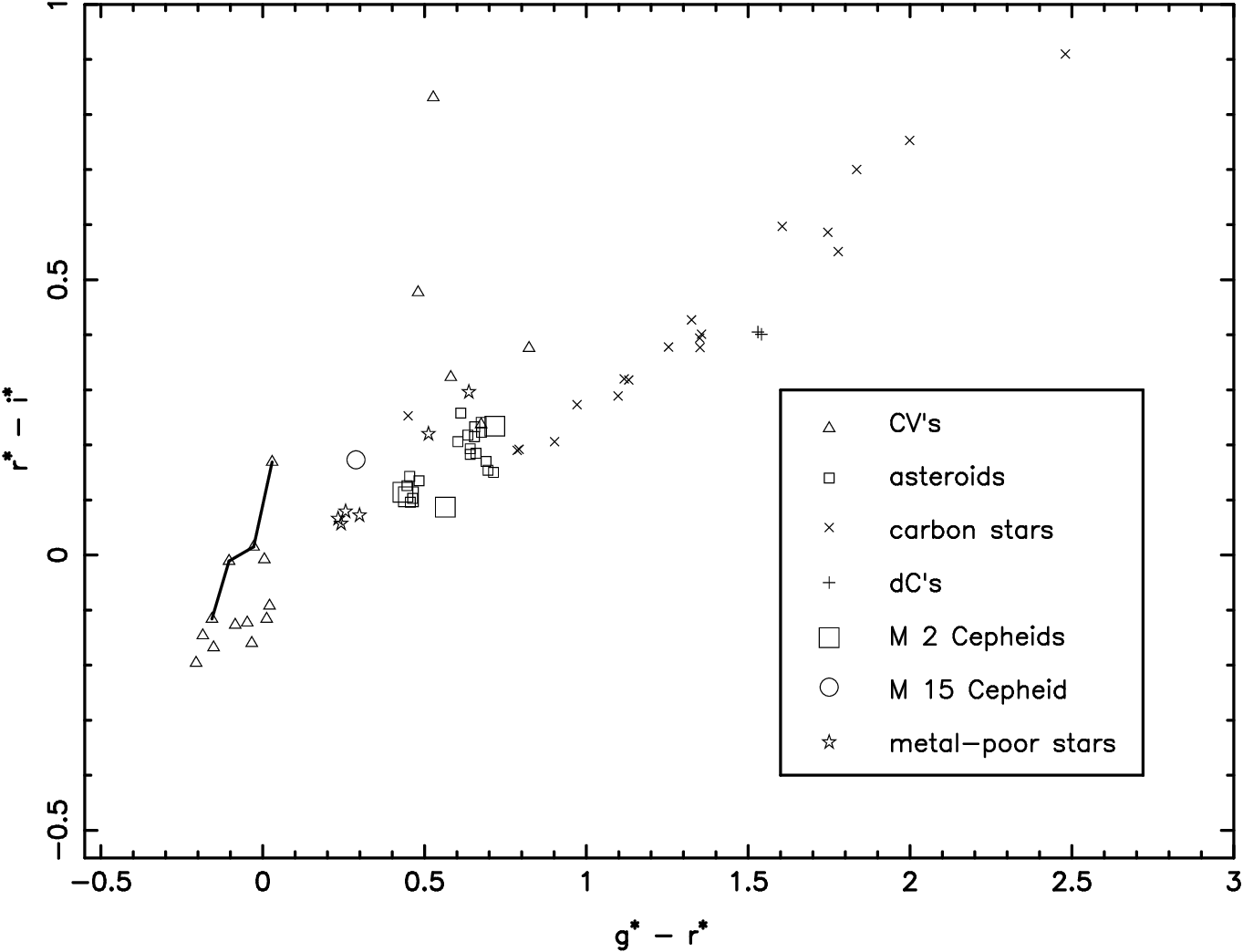,height=14cm,angle=-90}
\caption{r$^*-$i$^*$ vs. g$^*-$r$^*$ data for program objects
from Tables 2 through 5.  The triangles connected by 
a line are data of four separate nights for
RX Andromedae, which was recovering from an outburst.}
\end{figure*}

\begin{figure*}  
\psfig{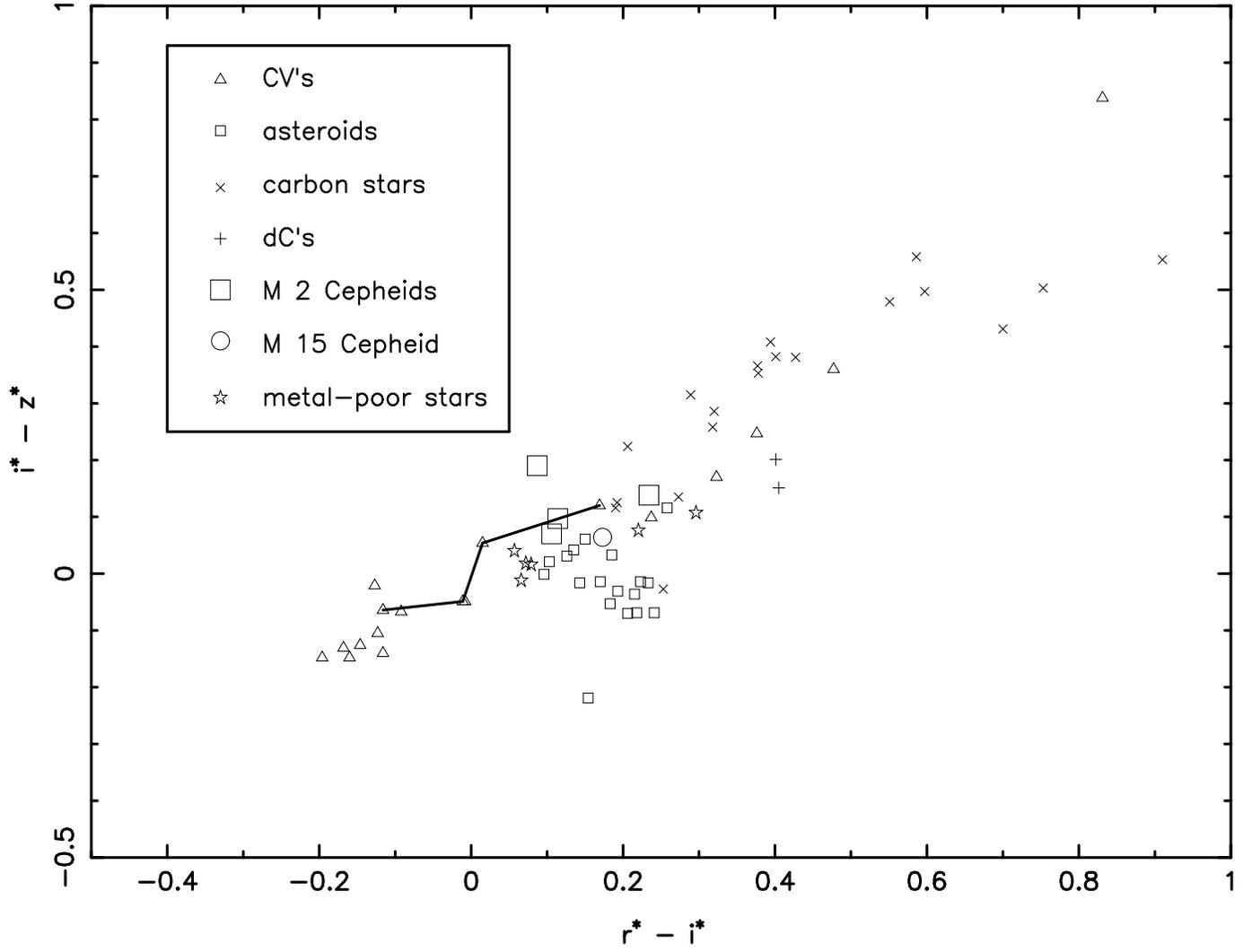}
\caption{i$^*-$z$^*$ vs. r$^*-$i$^*$ data for program objects
from Tables 2 through 5.  The triangles connected by 
a line are data of four separate nights for
RX Andromedae, which was recovering from an outburst.}
\end{figure*}

\begin{figure*}  
\psfig{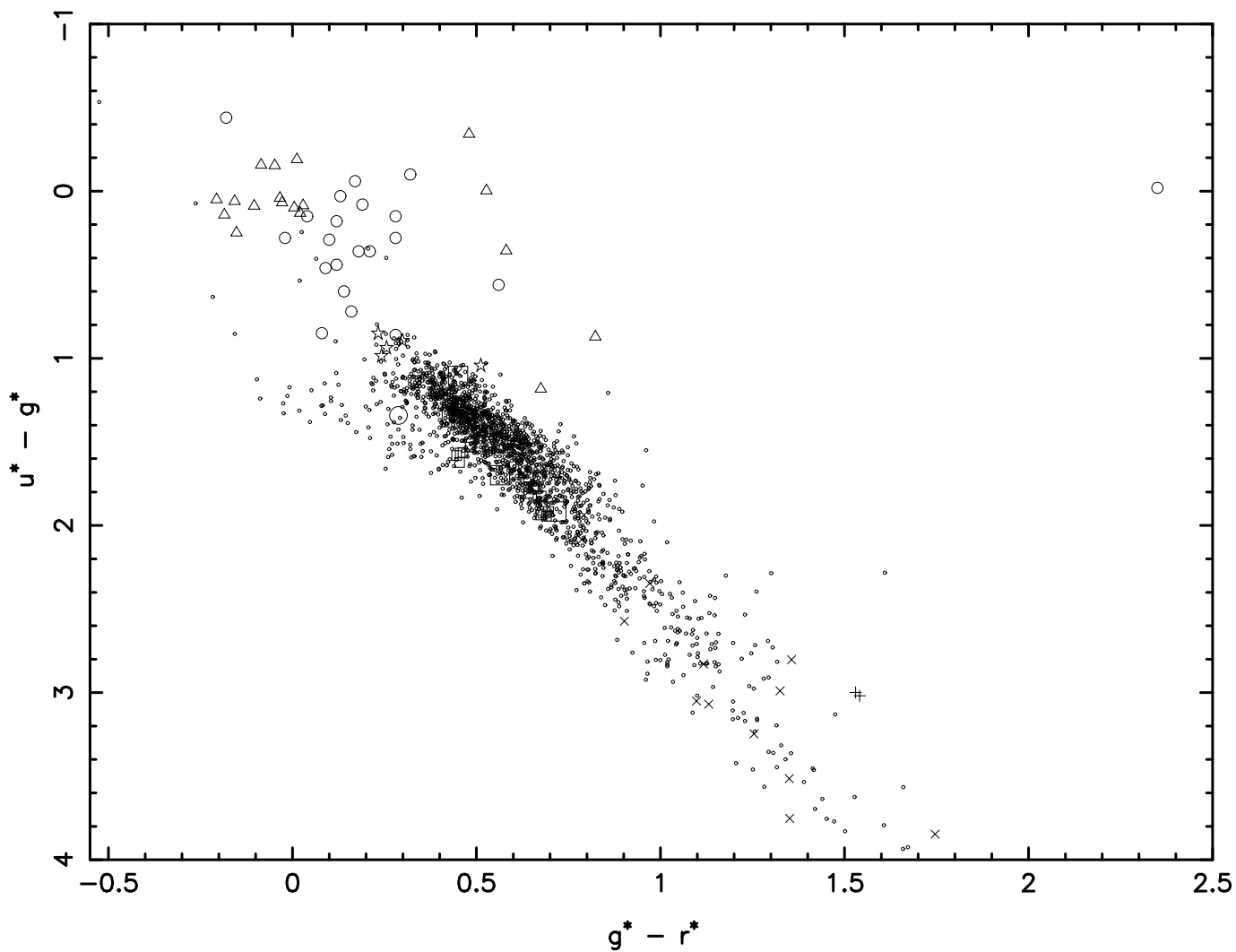}
\caption{u$^*-$g$^*$ vs. g$^*-$r$^*$
data for program objects, quasars, and the stellar locus obtained
from roughly 80 fields at a variety of galactic latitudes.
Symbols are the same as in Figs. 2 to 4, plus these
additional ones: tiny dots represent field stars,
and smaller open circles represent quasars 
from Richards et al. (1997), with redshifts ranging from
0.06 to 2.72. The three carbon stars reddest in
u$^*-$g$^*$ are off the bottom of the plot. Only 2 field
stars are observed to have u$^*-$g$^* > 4.0$.}
\end{figure*}

\begin{figure*}  
\psfig{figure=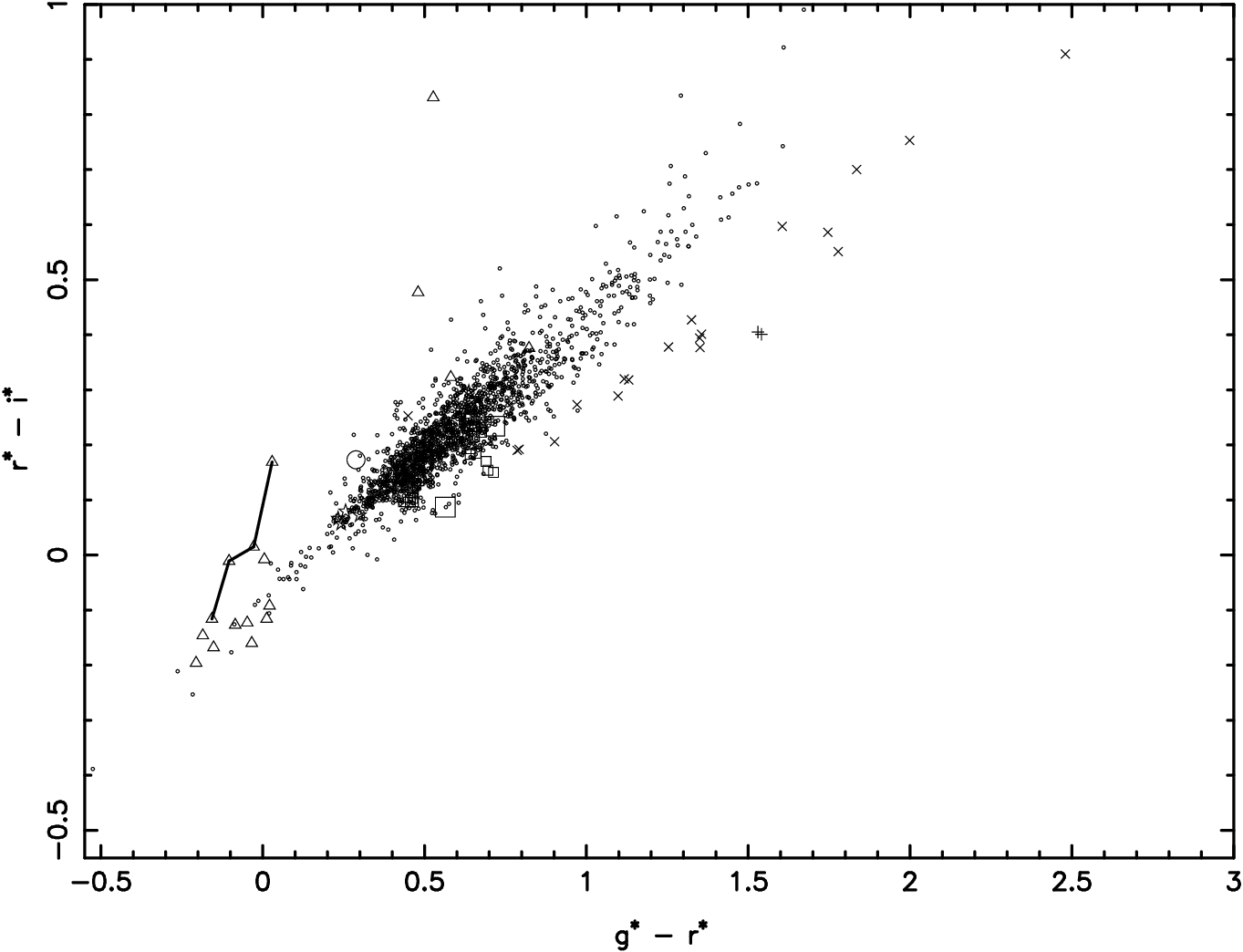,height=14cm,angle=-90}
\caption{r$^*-$i$^*$ vs. g$^*-$r$^*$ data for program objects and
field stars.  Symbols are the same as in Fig. 5.}  
\end{figure*}

\begin{figure*}  
\psfig{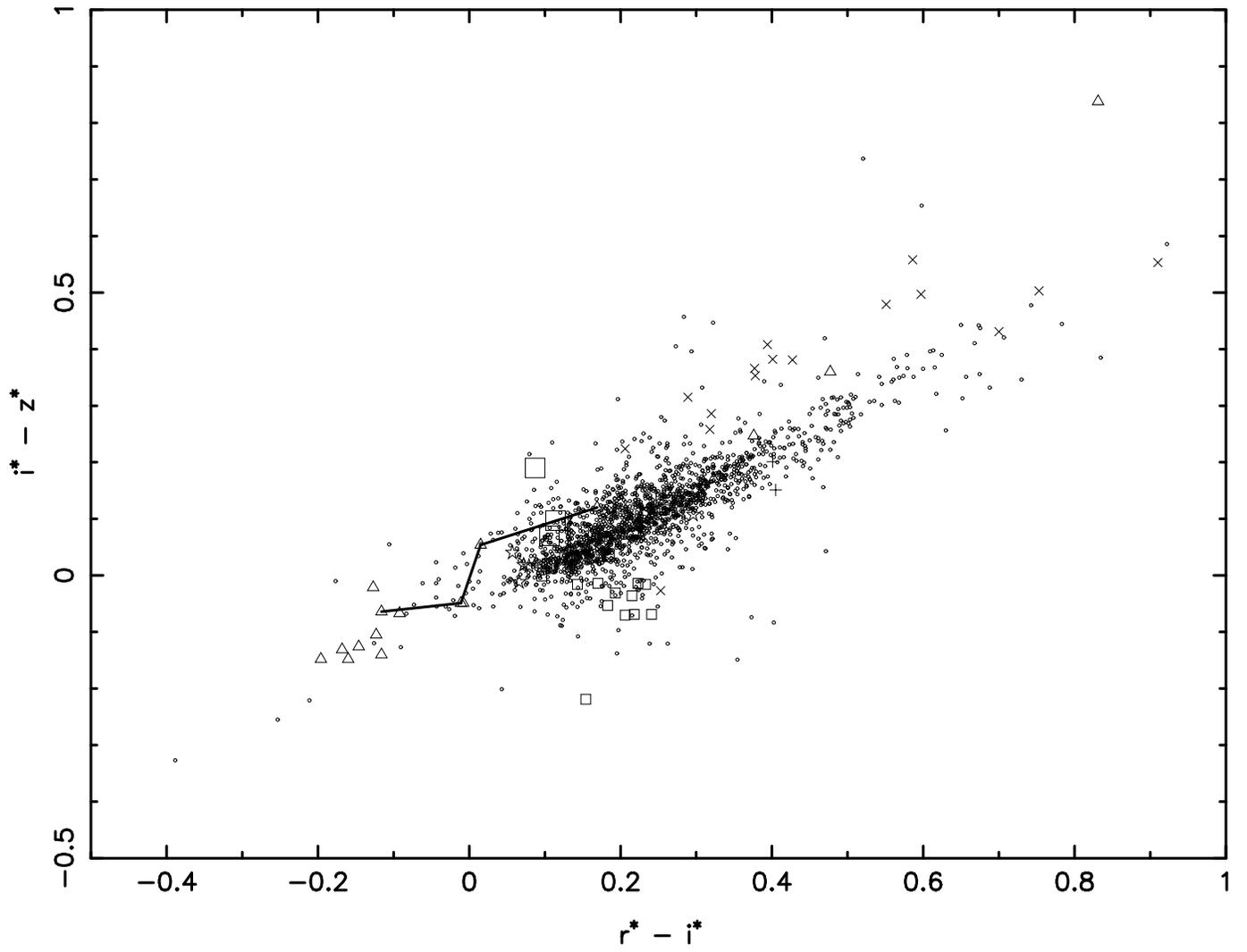}
\caption{i$^*-$z$^*$ vs. r$^*-$i$^*$ data for program objects
and field stars.  Symbols are the same as in Fig. 5.}  
\end{figure*}

\begin{figure*}  
\psfig{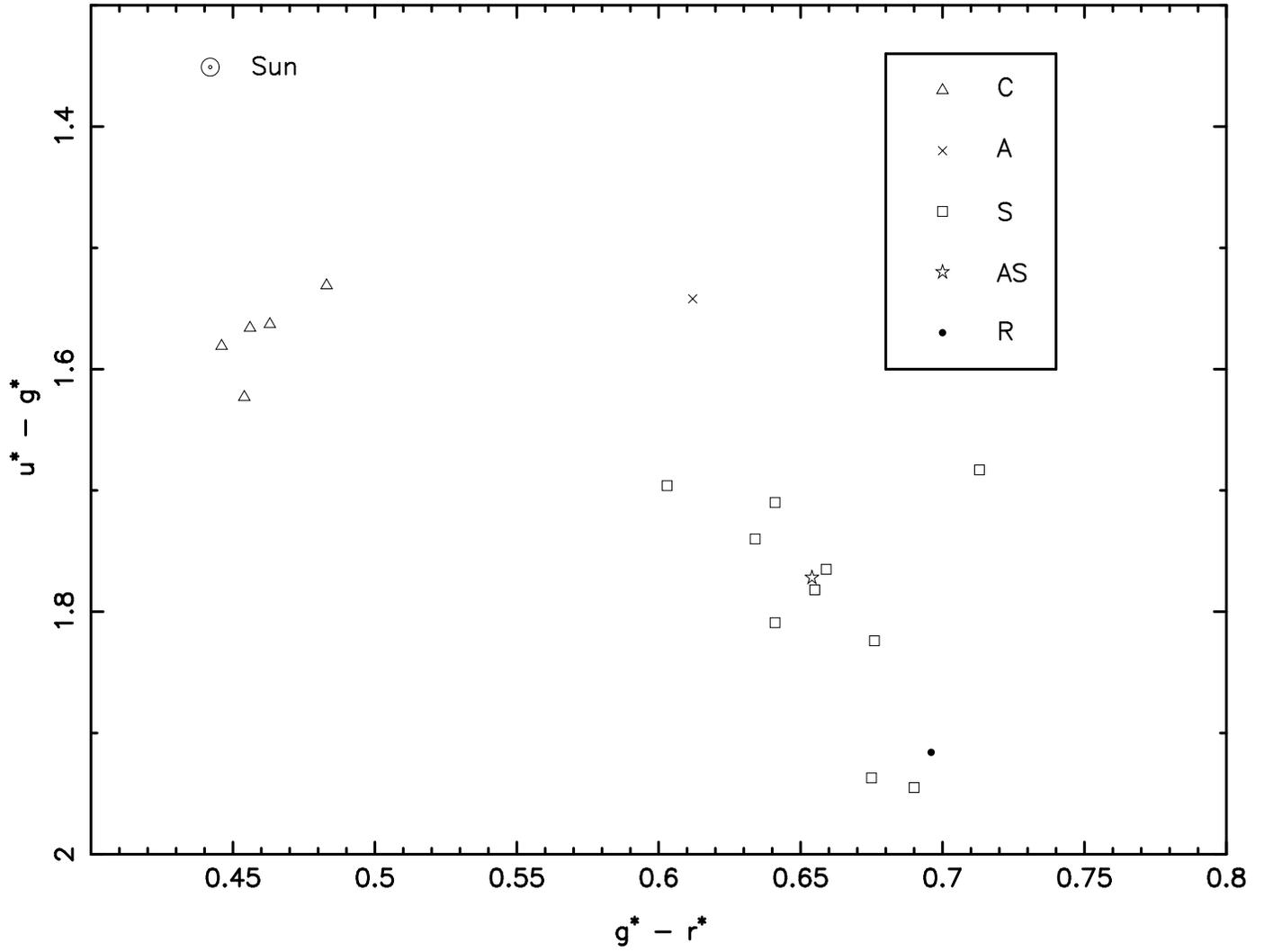}
\caption{u$^*-$g$^*$ vs. g$^*-$r$^*$ data for asteroids.
Three asteroids were measured on more than one night.  We plot 
all nightly means.}
\end{figure*}

\begin{figure*}  
\psfig{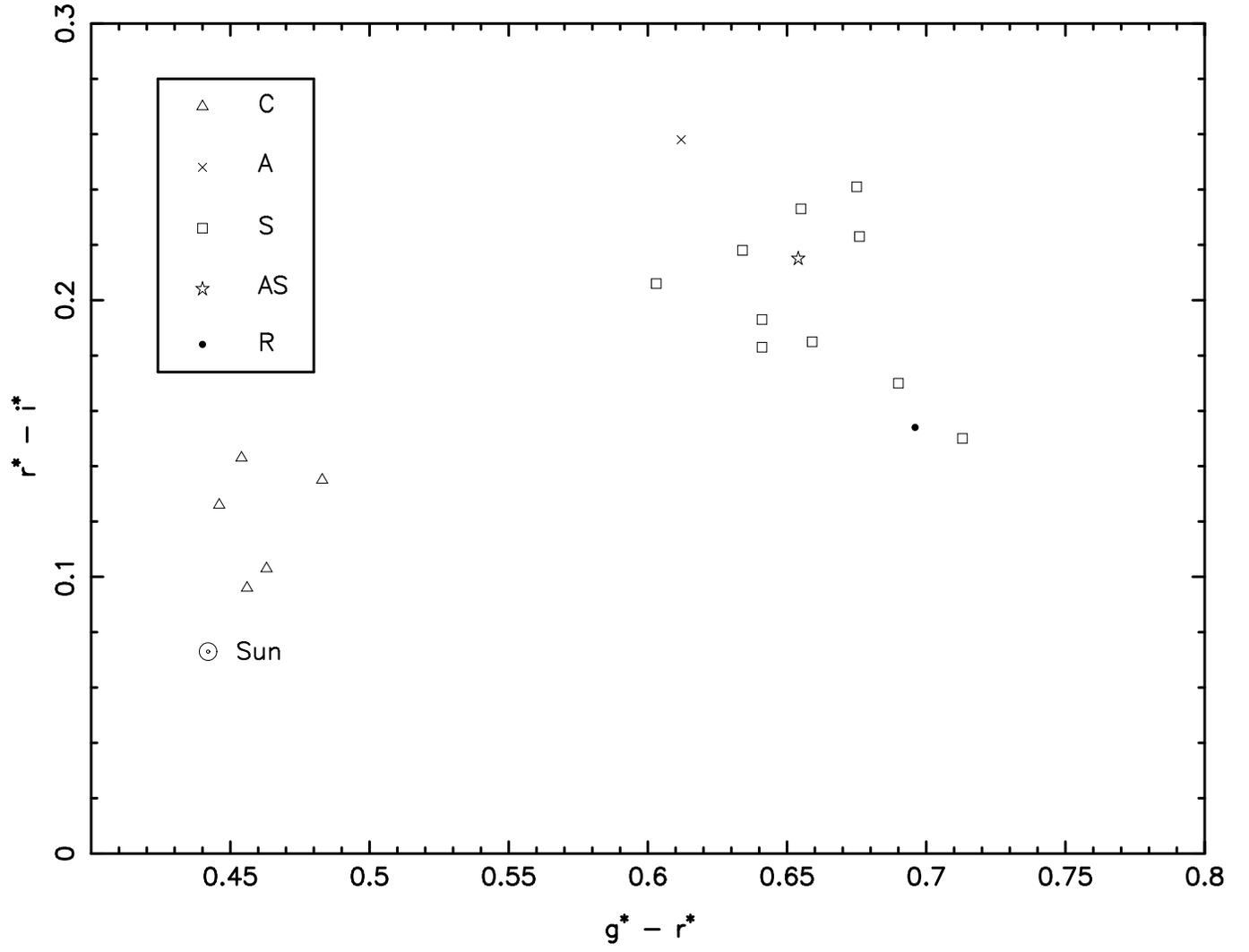}
\caption{r$^*-$i$^*$ vs. g$^*-$r$^*$ data for asteroids.}
\end{figure*}

\begin{figure*}  
\psfig{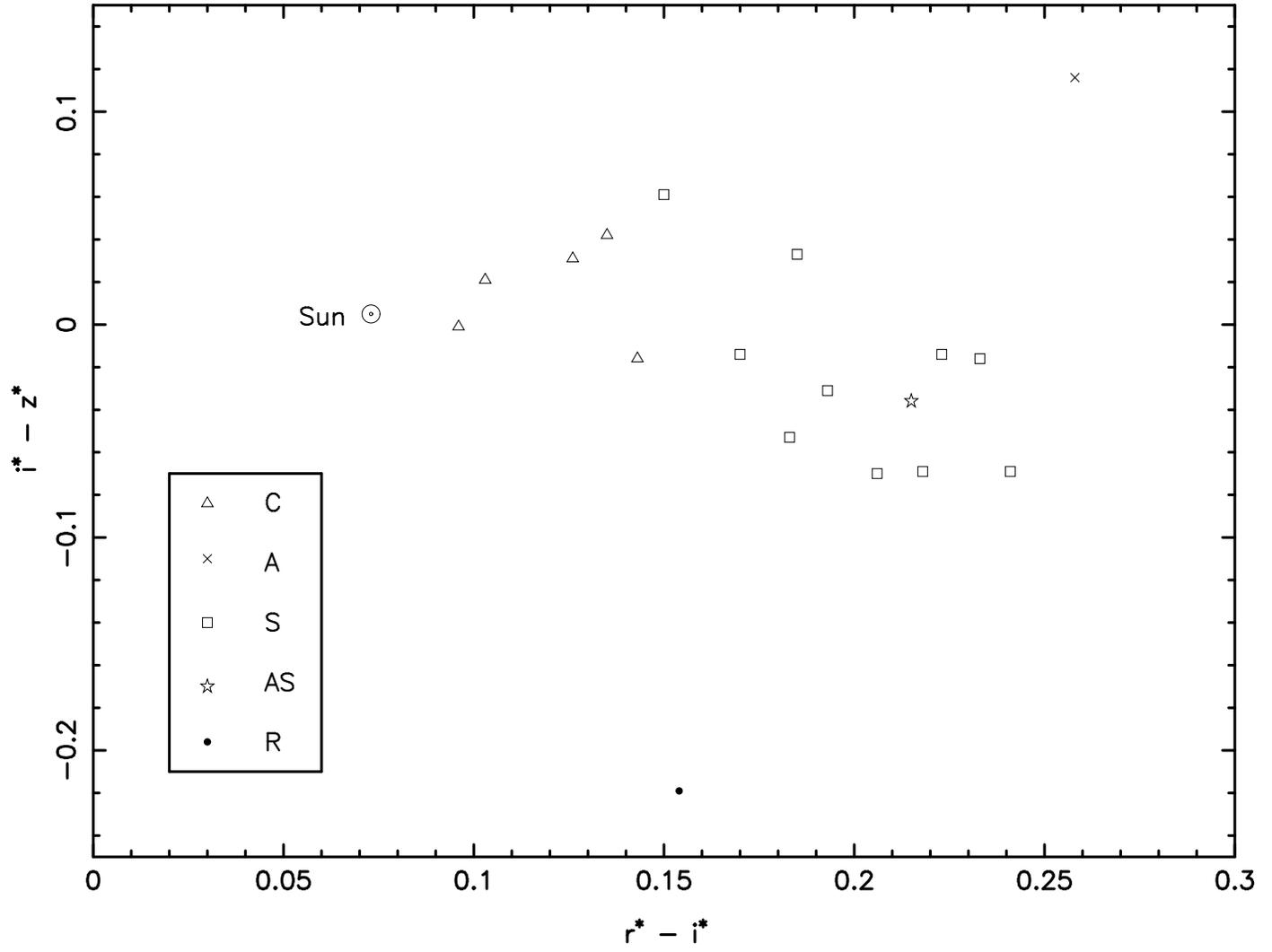}
\caption{i$^*-$z$^*$ vs. r$^*-$i$^*$ data for asteroids.}
\end{figure*}

\begin{figure*}  
\psfig{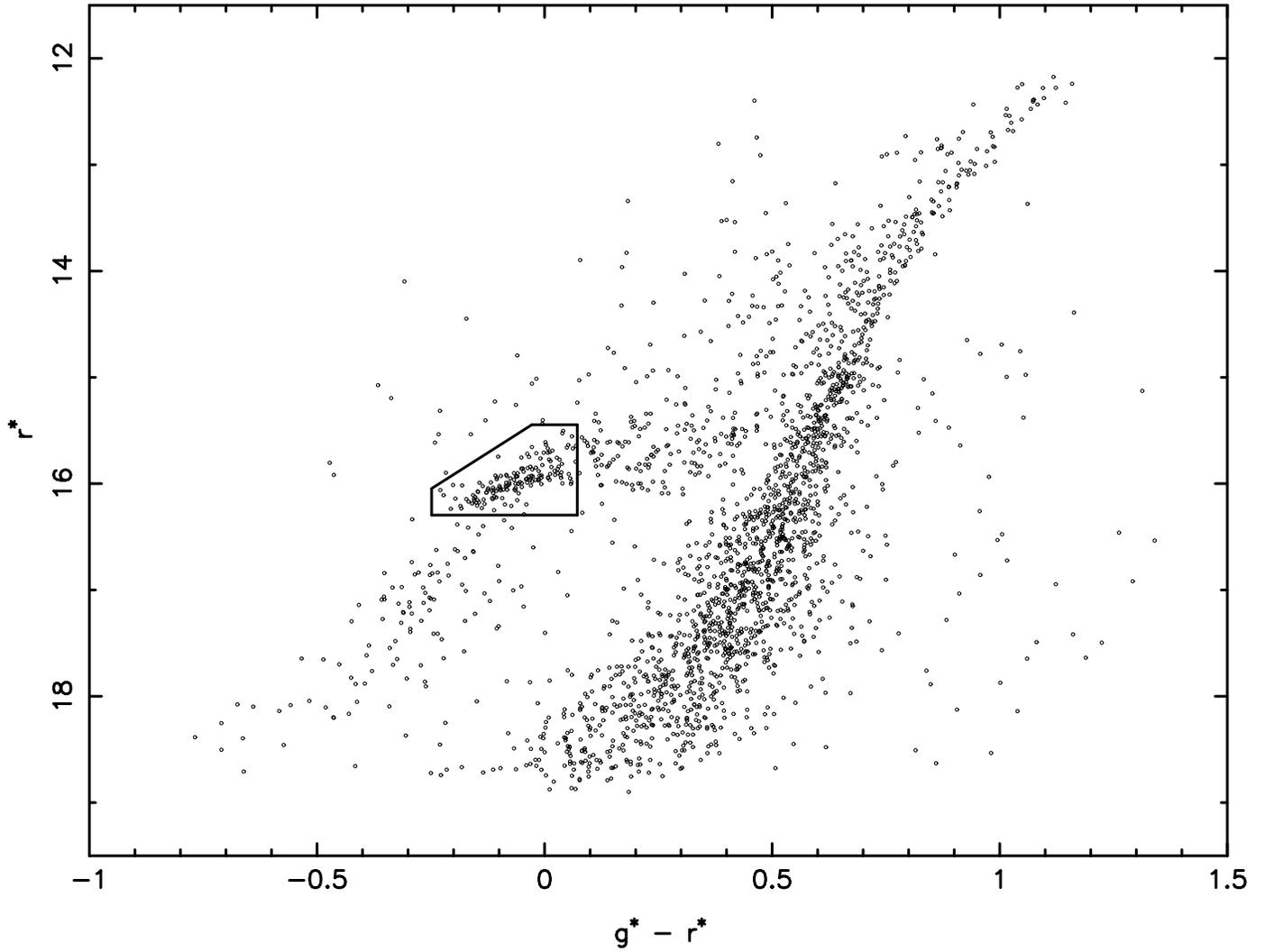}
\caption{Color-magnitude diagram for the globular cluster M 15.
The blue horizontal branch stars are located in the pentagonal
box.  Some RR Lyr stars are found in this box, but most are
found in a region of comparable size just to the right 
of the box, and slightly higher.}
\end{figure*}

\begin{figure*}  
\psfig{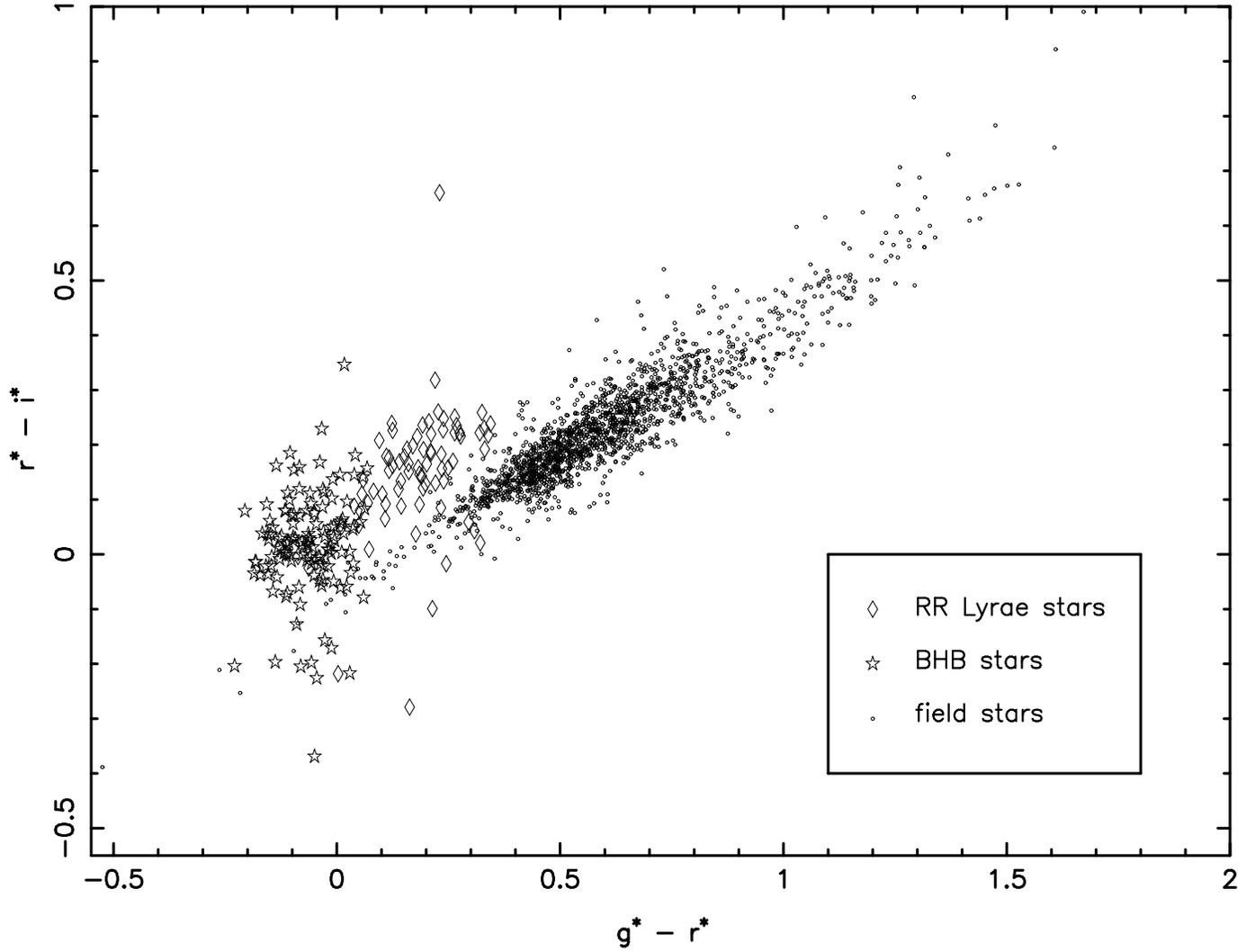}
\caption{r$^*-$i$^*$ vs. g$^*-$r$^*$ data for M 15 stars.
RR Lyr stars were identified using the coordinates
given by Sawyer Hogg (1973). Outliers
are most likely due to the effect of crowding discussed in the text.}
\end{figure*}

\begin{figure*}  
\psfig{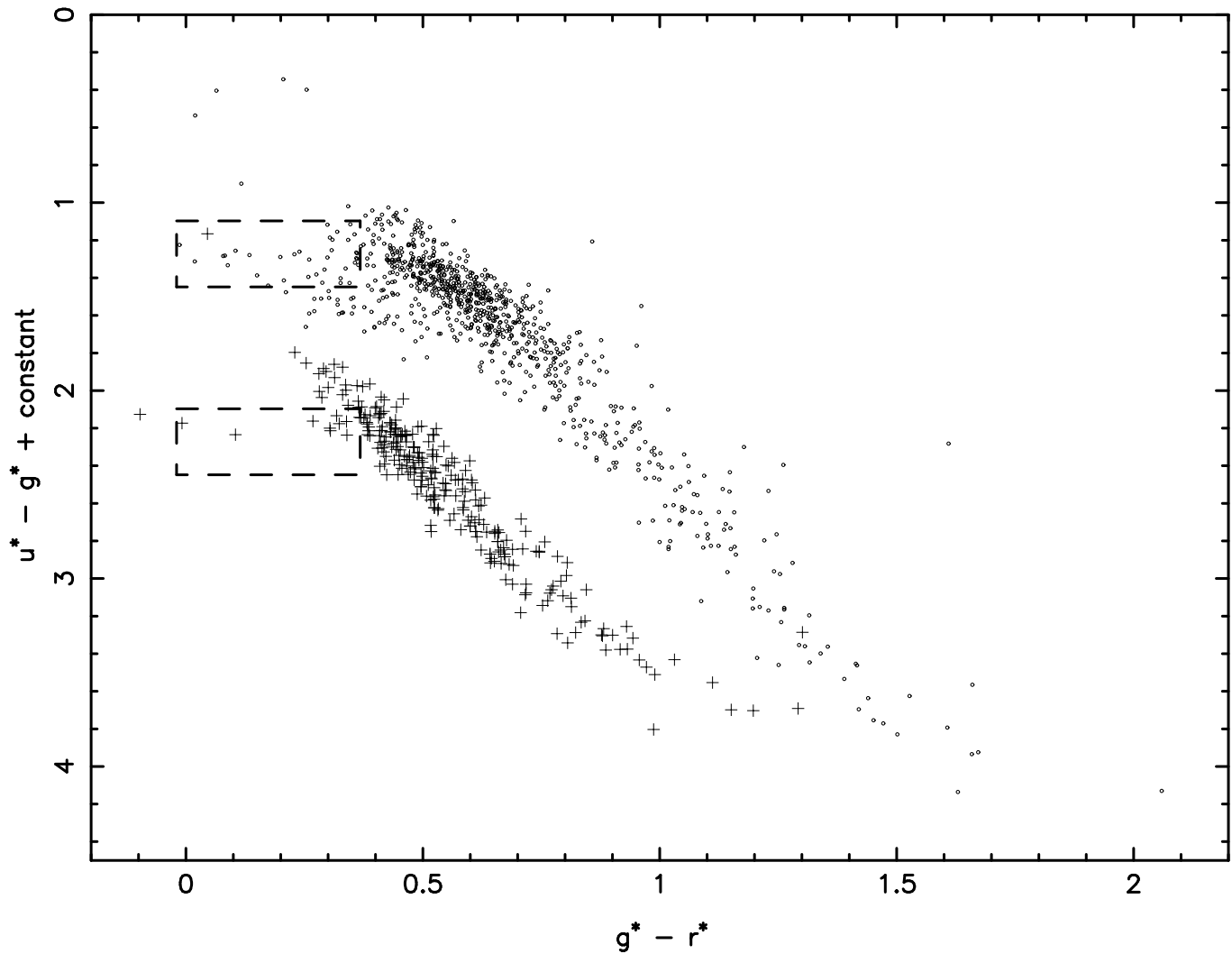}
\caption{u$^*-$g$^*$ vs. g$^*-$r$^*$ stellar locus for field stars
with galactic latitude 0$^o < |b| < 19^o$ (dots) and with
42$^o < |b| < 90^o$ (pluses).  The lower set of points has been
offset by +1.00 mag in u$^*-$g$^*$ for the purpose of display.  
The boxes indicate the
location of RR Lyr stars in M~15 ($\pm 2\sigma$ from mean value in
each direction). The BHB stars are located to the left of the boxes.}
\end{figure*}


\begin{references}
\reference{bf84} Berger, J., \& Fringant, A.-M. 1984, \aaps, 58, 565
\reference{b91} Bothun, G., Elias, J. H., MacAlpine, G., Matthews,
K., Mould, J. R., Neugebauer, G., \& Reid, I. N. 1991, \aj, 101, 2220
\reference{b65} Brownlee, K. A. 1965, Statistical Theory and Methodology
in Science and Engineering (New York: Wiley),  on p. 300
\reference{d94} Deutsch, E. W. 1994, \pasp, 106, 1134
\reference{f96} Fukugita, M., Ichikawa, T., Gunn, J. E.,
Doi, M., Shimasaku, K., \& Schneider, D. P. 1996, \aj, 111, 1748 (F96)
\reference{g98} Green, P. J. 1998, in The Carbon Star Phenomenon,
ed. R. F. Wing (Dordrecht: Kluwer), IAU Symposium 177, in press
\reference{gm94} Green, P. J., \& Margon, B. 1994, \apj, 423, 723
\reference{gk93} Gunn, J. E., \& Knapp, G. R. 1993, in Sky Surveys:
Protostars to Protogalaxies, ed. B. T. Soifer (San Francisco:
Astronomical Society of the Pacific), 267
\reference{gw95} Gunn, J. E., \& Weinberg, D. H. 1995, in
Wide Field Spectroscopy and the Distant Universe, eds. S. J. Maddox \&
A. Arag\'on-Salamanca (Singapore: World Scientific), pp. 3-14  
\reference{h62} Hardie, R. H. 1962, in Astronomical Techniques,
ed. W. A. Hiltner (Chicago: Univ. of Chicago Press), 178
\reference{h96} Harris, W. E. 1996, \aj, 112, 1487
\reference{gcvs} Kholopov, P. N. 1987, General Catalogue of Variable
Stars, 4th ed., vol. III (Moscow: Nauka)
\reference{i90} Isobe, T., Feigelson, E. D., Akritas, M. G., \&
Babu, G. J. 1990, \apj, 364, 104
\reference{k87} Krisciunas, K., et al. 1987, \pasp, 99, 887
\reference{l92} Landolt, A. U. 1992, \aj, 104, 340
\reference{l96} Lee, P., et al. 1996, Icarus 120, 87
\reference{l98} Lenz, D. D., Newberg, H. J., Rosner, R., Richards,
G. T., \& Stoughton, C. 1998, \apjs, 119, in press
\reference{m98} Margon, B. 1998, Phil. Trans. Roy. Soc. A, in press
(astro-ph/9805314)
\reference{bm84} Margon, B., Aaronson, M., Liebert, J., \& Monet, D.
1984, \aj, 89, 274
\reference{m96} Mateo, M. 1996, ``Canary Islands Winter School 1996:
Stellar Photometry Exercises Using DoPHOT''
\reference{new97} Newberg, H., Richards, G., Lenz, D., Fan, X.,
Richmond, M., \& Yanny, B. 1997, \baas, 29, 1385
\reference{r97} Richards, G. T., Yanny, B., Annis, J., Newberg,
H. J. M., McKay, T. A., York, D. G., \& Fan. X. 1997, \pasp,
109, 39
\reference{sp88} Sanduleak, N., \& Pesch, P. 1988, \apjs, 66, 387
\reference{hsh} Sawyer Hogg, H. 1973, Publ. David Dunlap Obs., 3, No. 6
\reference{dophot} Schechter, P., Mateo, M., \& Saha, A. 1993,
\pasp, 105, 1342
\reference{ss95} Silbermann, N. A. \& Smith, H. A. 1995, \aj, 110, 704
\reference{skb69} Slettebak, A., Keenan, P. C., \& Brundage, R. K. 1969,
\aj, 74, 373
\reference{tb89} Tholen, D. J., \& Barucci, M. A. 1989, in Asteroids II,
eds. R. P. Binzel, T. Gehrels, \& M. S. Matthews, (Tucson:
Univ. of Arizona Press), 298
\reference{ti98} Totten, E. J., \& Irwin, M. J. 1998, \mnras, 294, 1
\reference{w95} Warner, B. 1995, Cataclysmic Variable Stars (Cambridge:
Cambridge Univ. Press)
\reference{x95} Xu, S., Binzel, R. P., Burbine, T. H., \& Bus, S. J.
1995, Icarus, 115, 1
\end{references}
\end{document}